\documentclass[11pt]{article}
\usepackage{graphicx}
\usepackage{graphics}
\usepackage{dsfont}

\newcommand{\HirzeF}{\mathds{F}}

\usepackage{epsfig}
\usepackage{amsmath,amssymb,amsthm,amscd}

\def\ZZ{{\mathbb Z}}

\def\2{{1\over2}}
\let\<=\langle \let\>=\rangle
\def\new#1\endnew{{\bf #1}}
\def\ifundefined#1{\expandafter\ifx\csname#1\endcsname\relax}
\ifundefined{draftmode}\else    \input draftmode        \fi
\let\Msize=\footnotesize             
\def\BM{\Msize\begin{matrix}}           \def\EM{\end{matrix}}
\def\MN M:#1 #2 N:#3 #4 {{(#1_{#2},#3_{#4})}}
\def\MNH M:#1 #2 N:#3 #4 H:#5,#6 [#7]{{(#1_{#2},#3_{#4})^{#5,#6}_{#7}}}

\newcommand{\ds}{\displaystyle}
\newcommand{\cF}{{\cal F}}

\def\dd{\mathrm{d}}

\newcommand{\half}{{1\over 2}}

\newcommand{\be}{\begin{equation}}
\newcommand{\ee}{\end{equation}}
\newcommand{\bea}{\begin{eqnarray}}
\newcommand{\eea}{\end{eqnarray}}

\def\half{\frac{1}{2}}

\begin{document}

%%%%%%%%%%%%%%%%%
\begin{titlepage}
{}~ \hfill\vbox{ \hbox{} }\break

\rightline{Bonn-TH-10-06}
\rightline{IPMU10-0153} \vskip 1cm

\centerline{\Large \bf Direct integration for general $\Omega$ backgrounds}  \vskip 0.5 cm
\renewcommand{\thefootnote}{\fnsymbol{footnote}}
\vskip 30pt \centerline{ {\large \rm Min-xin Huang
\footnote{minxin.huang@ipmu.jp} and Albrecht Klemm
\footnote{aklemm@th.physik.uni-bonn.de} }} \vskip .5cm \vskip 30pt
\vspace*{4.0ex}
\begin{center}
{\em $^\star$ Institute for the Physics and Mathematics of the Universe (IPMU),  \\ University of Tokyo, Kashiwa, Chiba 277-8582, Japan}\\ [3 mm]
{\em $^\dagger$\ \ Physikalisches Institut, Universit\"at Bonn, 
            D-53115 Bonn, FRG}
\end{center}

\setcounter{footnote}{0}
\renewcommand{\thefootnote}{\arabic{footnote}}
\vskip 60pt
\begin{abstract}
We extend the direct integration method of the holomorphic anomaly
equations to general $\Omega$ backgrounds $\epsilon_1\neq -\epsilon_2$ 
for pure SU(2) N=2 Super-Yang-Mills theory and topological string 
theory on non-compact Calabi-Yau threefolds. We find that an  
extension of the holomorphic anomaly equation, modularity  and 
boundary conditions provided by the perturbative terms as well as by 
the gap condition at the conifold are sufficient to solve the generalized 
theory in the above cases. In particular we use the method to solve 
the topological string for the general $\Omega$ backgrounds on 
non-compact toric Calabi-Yau spaces. The conifold boundary 
condition follows from  that the N=2 Schwinger-Loop calculation 
with BPS states coupled to a self-dual and an anti-self-dual 
field strength. We calculate such BPS states also for the 
decompactification limit of Calabi-Yau spaces with 
regular K3 fibrations and  half K3s embedded in Calabi-Yau backgrounds.
\end{abstract}
\end{titlepage}
\vfill \eject

%%%%%%%%%%%%%%%%%%%%%%%%%%%%%%%%%%%%%%%%%%%%%%%%%%%%%%%%%%%%%

\newpage

\baselineskip=16pt

\tableofcontents

\section{Introduction}

Nekrasov's instanton  calculations for the N=2 
supersymmetric gauge theory~\cite{Nekrasov} completes 
the program of~\cite{LNS} and confirms the Seiberg-Witten 
prepotential as the leading contribution in the 
asymptotically free region from the microscopic 
field theory perspective. These instanton calculations 
have been made mathematically more rigorous in~\cite{NO, NY1}. 
Higher order contributions in Nekrasov partition function correspond 
to gravitational couplings of the gauge theory, and are organized by a
topological genus expansion.  
The genus one formula is also mathematically proven 
in~\cite{NY2}. 
In previous works, we computed the higher genus terms in $SU(2)$ 
Seiberg-Witten theory (with fundamental matter)~\cite{HK1} 
(\cite{HK2}) in terms of generators  of modular forms  w.r.t.  
the monodromy group, which is a  subgroup of $SL(2,Z)$, 
using holomorphic anomaly  equations~\cite{BCOV} and novel 
boundary conditions at the special points of the moduli space. 
Our formulae constitute well-defined mathematical conjectures 
that sum up all instanton contribution of Nekrasov's 
partition function at  fixed genus in a closed from, which defines 
it explicitly at every point on the Coulomb branch. 

There are two deformation parameters $\epsilon_1, \epsilon_2$ in Nekrasov 
partition function. Our higher genus formulae in~\cite{HK1, HK2} 
correspond to the case $\epsilon_1=-\epsilon_2$  or 
$\beta:=-\frac{\epsilon_1}{\epsilon_2}=1$, where the technique 
of holomorphic anomaly equations from topological string 
theory is applicable. Recently, it has become an interesting 
question to study the general case of arbitrary $\beta$-backgrounds 
due to several developments. 

Firstly, the AGT (Alday-Gaiotto-Tachikawa) conjecture~\cite{AGT} 
relates the Nekrasov function at general deformation parameter 
(at a fixed instanton number) to correlation functions in Liouville 
theory. A matrix model with a modified measure, the called 
$\beta$-ensemble~\cite{Dijkgraaf:2009pc}, was related to the 
general $\beta$-deformations of gauge theories.

Secondly, in the BPS interpretation of the topological string 
partition function, there is a natural meaning of the $\epsilon_1, 
\epsilon_2$ expansion. It gives refined information about the 
cohomology of the moduli space of the BPS states, while  the 
$\epsilon_1=-\epsilon_2$ slice computes complex structure 
invariant indices. A refined topological vertex was proposed 
in~\cite{IKV} that generalizes the topological string partition 
function for non-compact toric Calabi-Yau manifolds, which have 
no complex structure deformations. It was shown to reduce to 
the Nekrasov partition function for general deformation parameters 
$\epsilon_1, \epsilon_2$ in the field theory limit. 

In this paper we describe first the B-model approach of direct integration 
for the simplest deformed N=2 gauge theory in section~\ref{SWtheory}. 
Similar results have been also obtained recently, in fact 
in more generality in~\cite{Krefl:2010fm}. It is 
clear already  from the perturbative test of~\cite{Mironov:2010ym}
that the $\beta$-ensemble for the matrix models 
associated to Seiberg-Witten theories suggested 
in~\cite{Dijkgraaf:2009pc} leads to the possibility to remodel 
the B-model along the line of~\cite{BKMP} from the spectral 
curve using the formalism of~\cite{Eynard:2007kz}.

A direct implementation of the deformed $\beta$-ensemble to the matrix models 
associated to  the topological string on local 
Calabi-Yau spaces seems not straightforward.
We found that even the $\beta$-ensemble for the Chern-Simons 
matrix model, describing the resolved conifold in the canonical 
parameterization, fails\footnote{We thank Marcos Mari\~no 
for sharing insights into similar attempts.} to reproduce the 
known results~\cite{IKV} for this geometry if 
$\epsilon_1\neq - \epsilon_2$.  We therefore move on to calculate 
general deformations in topological string theory in 
section~\ref{refinedtopstring}.

In section~\ref{BPSamplitude} we explain first the interpretation 
of the refined topological string expansion in terms of the 
cohomology of the moduli space of BPS states. The BPS picture 
yields the generalized gap condition at the conifold locus 
in section~\ref{leadingcoefficients} and the large radius 
conifold expansion in section~\ref{resconifold}.  
 
We then make predictions for the generalized BPS invariants in the 
decompactification limit of $K3$ fibered Calabi-Yau spaces for 
large base space using heterotic type II duality in section~\ref{het/typeII}  
and analyze in section~\ref{halfk3} a similar setting for the half $K3$, 
which by a T-duality predicts a sector of the partition function 
for compactifications of $N=4$ SYM on manifolds with $b_{+}=1$.

Using the generalized holomorphic anomaly equation and boundary 
condition we extend the methods of~\cite{HK1, HK2,HKR}.  
I.e. we perform the direct integration for general $\beta$-backgrounds 
on local toric Calabi-Yau spaces using the generalized gap condition 
discussed in~\ref{leadingcoefficients}.  We then consider  
non-compact manifolds, such as ${\cal O}(K_B)\rightarrow B$, 
where the base $B$ is a toric manifold. As examples we present 
the cases $B=\mathbb{P}^2$ and $B=\mathbb{P}^1\times \mathbb{P}^1$ 
in sections~\ref{P1P1} and \ref{P2} respectively. We hope that 
our analytic expressions for the amplitudes will help to find 
a matrix model description.

\section{Seiberg-Witten gauge theory}
\label{SWtheory}
The Nekrasov partition function consists of the perturbative 
contributions and the instanton contributions
\begin{equation}
Z(a,\epsilon_1,\epsilon_2)=Z_{pert} (a,\epsilon_1,\epsilon_2) Z_{inst} (a,\epsilon_1,\epsilon_2)
\end{equation}
In this paper we consider only the pure $SU(2)$ case, so there is only one
Seiberg-Witten period $a$ and we choose the cut-off parameter  in~\cite{NO} 
to be $\Lambda=1/16$ which can be recovered by dimensional analysis in the
formula. So the function essentially depends on 3 parameters 
$a, \epsilon_1, \epsilon_2$. The logarithm of the Nekrasov function can be expanded as 
\begin{equation}
\log Z(a,\epsilon_1,\epsilon_2)=  \sum_{i,j=0}^\infty (\epsilon_1+\epsilon_2)^{i}(\epsilon_1\epsilon_2)^{j-1}F^{(\frac{i}{2},j)}(a)
\label{mixedexpansion}
\end{equation}
Then the genus zero $F^{(0,0)}$ is the prepotential well known from the work of Seiberg-Witten \cite{SW1}, and the formula $F^g$ with $g>1 $ for the case of $\epsilon_1+\epsilon_2=0$ 
in~\cite{HK1, HK2} correspond to $F^g=F^{(0,g)}$ in our current notation. 
It turns out that for the models that we will study, when $i$ is an odd integer, $F^{(\frac{i}{2},j)}(a)$
vanishes except a trivial term from the perturbative contributions. 
So we will only need to consider $F^{(i,j)}(a)$ with $i,j$ non-negative 
integers. This is not always true for all models. In particular, the
$F^{(\frac{1}{2},0)}$ is non-vanishing for $SU(2)$ Seiberg-Witten theory 
with $N_f=1$ massless flavor. For this interesting case, 
as well as the massless $N_f=2,3$ theories, see 
the recent paper~\cite{Krefl:2010fm}. 

\subsection{Generalized holomorphic anomaly equations}

It turns out that the topological amplitudes $F^{(g_1,g_2)}$ satisfy for
$g_1+g_2\geq 2$ a generalized holomorphic anomaly equation
\begin{eqnarray} \label{generalizedBCOV00}
\bar{\partial}_{\bar{i}} F^{(g_1,g_2)}= \frac{1}{2}\bar{C}_{\bar{i}}^{jk}\big{(}D_jD_kF^{(g_1,g_2-1)}
+{\sum_{r_1,r_2} }^{\prime}  D_jF^{(r_1,r_2)}D_kF^{(g_1-r_1,g_2-r_2)}\big{)}
\end{eqnarray}
where the prime denotes that the sum over $r_1,r_2$ does not include
$(r_1,r_2)=0$ and $(r_1,r_2)=(g_1,g_2)$, 
and the first term on the right hand side is understood to be zero 
if $g_2=0$. This equation reduces to the ordinary BCOV holomorphic 
anomaly equation when $g_1=0$, and is a simplification of the 
extended holomorphic anomaly equation in~\cite{Krefl:2010fm} 
without the so called Griffiths infinitesimal invariant, 
which turns out to be vanishing for the models we study. 

To integrate the holomorphic anomaly equation and write out the compact expressions for the higher genus amplitudes, we first express the Seiberg-Witten period $a$ and  Coulomb modulus $u$ in terms of modular functions of the coupling 
$\tau\sim \frac{1}{2\pi i} \frac{\partial^2 F^{(0,0)}}{\partial a^2}$ as  
\begin{eqnarray}
a = \frac{E_2(\tau)+\theta_3^4(\tau)+\theta_4^4(\tau)}{3\theta_2^2(\tau)}, ~~~
u = \frac{\theta_3^4(\tau) +\theta_4^4(\tau) }{\theta_2^4(\tau)} \label{autau}
\end{eqnarray}
In the cusp limit $\tau\rightarrow i\infty$, we find $q=e^{2\pi i \tau} \rightarrow 0$ and $a\sim q^{-\frac{1}{4}}$, $u\sim q^{-\frac{1}{2}}$. We can express $a,u, q$ in terms of series expansion of each other, from the above relations and the well known series expansion formulae of the Theta functions and Eisenstein series.  It is proven in \cite{NY2} that the genus one formulae for Nekrasov function are 
\begin{equation} \label{genus1(2.6)}
F^{(0,1)}= -\log (\eta(\tau)), ~~~~  ÊF^{(1,0)}= -\frac{1}{6}\log (\frac{\theta_2^2}{\theta_3 \theta_4})
\end{equation}
 
The anholomorphic generator in the topological amplitudes is the shifted
Eisenstein series $\hat{E}_2=E_2(\tau)+\frac{6i}{\pi(\bar{\tau}-\tau)}$.  
Using some well-known results about the three-point coupling and the 
relations between parameters $a,u, \tau$ in (\ref{autau}), we find that 
(\ref{generalizedBCOV00}) becomes 
\begin{equation} \label{generalizedBCOV}
48 \frac{\partial F^{(g_1,g_2)}}{\partial E_2}= \frac{d ^2}{d a^2}F^{(g_1,g_2-1)} 
+(\sum_{r_1=0}^{g_1}\sum_{r_2=0}^{g_2})^\prime  (\frac{d F^{(r_1,r_2)}}{d a} ) ( \frac{d F^{(g_1-r_1,g_2-r_2)}}{d a} )
\end{equation}
If the above generalized holomorphic anomaly equation is true, then it will
determine $F^{(g_1,g_2)}$ recursively up to a rational function of modulus $u$
with a pole at the discriminant of Seiberg-Witten curve $u^2-1$ of degree $2(g_1+g_2)-2$.

The equation (\ref{generalizedBCOV}) applies to the case of $g_1+g_2\geq 2$. 
At genus one, we note that $F^{(0,1)}$ satisfies the ordinary BCOV 
holomorphic anomaly equation after we pass to the usual modular but
an-holomorphic completion of $\eta\rightarrow \sqrt{{\rm Im}(\tau)}|\eta(\tau)|^2$. 

As for $F^{(1,0)}$, we write 
\begin{eqnarray}
F^{(1,0)}= -\frac{1}{6}\log (\frac{\theta_2^2}{\theta_3 \theta_4})=\frac{1}{24}\log(u^2-1) 
\end{eqnarray}
We see that $F^{(1,0)}$ has only a logarithmic cut at the discriminant
$u^2-1$. It is already modular, needs no an-holomorphic modular 
completion and has therefore no holomorphic anomaly.

\subsection{Higher genus formulae and the dual expansion}

Problems associated to Riemann surfaces ${\cal C}_1$ of 
genus one such as $SU(2)$ N=2 SYM theories, topological 
string related to local del Pezzo surfaces of cubic 
matrix models have only one an-holomorphic 
generator $\hat E_2$ in the ring of modular 
objects generating all $F^{(g_1,g_2)}$. It is 
convenient to define an an-holomorphic generator 
of weight zero, e.g. $X=\hat E_2/\theta_2^4$ in the case 
above.   

In cases with one an-holomorphic generator $X$ 
the direct integration of the generalized holomorphic 
anomaly equation of the type (\ref{generalizedBCOV}) 
leads to the following general form of the $F^{(g_1,g_2)}$
\begin{equation} 
F^{(g_1,g_2)}=\frac{1}{\Delta^{2(g_1+g_2-1)}({u})} 
\sum_{k=0}^{3 g_2-3-g_1} X^k c^{(g_1,g_2)}_k({u})\ , 
\label{generalformFmn}
\end{equation}
where $\Delta({u})$ is the conifold 
discriminant of ${\cal C}_1$ and ${u}$ 
are holomorphic monodromy invariant parameters. 
All $c^{(g_1,g_2)}_i({u})$ are polynomial in these 
parameters. The extension of the generalized anomaly equations 
and the general form (\ref{generalformFmn}) to cases 
with more an-holomorphic generators $X_{ij}$ for theories 
related to Riemann surfaces ${\cal C}_{g>1}$  works 
along the lines discussed in~\cite{Aganagic:2006wq}
\cite{Grimm:2007tm}\cite{AL}.

In (\ref{generalformFmn}) all $c^{(g_1,g_2)}_{i>0}({ u})$ are 
determined by the generalized holomorphic anomaly equation, while 
the holomorphic ambiguity $c^{(g_1,g_2)}_{0}({ u})$ 
must be determined from the boundary conditions. We find 
that the expansion at the conifold divisor in the moduli space 
and in particular the gap condition in this expansion together 
with regularity at other limits in the moduli space 
and the knowledge of the classical terms are sufficient 
to completely fix $c^{(g_1,g_2)}_{0}({u})$.  Note that regularity of  
$F^{(g_1,g_2)}$ the $u\rightarrow \infty $ limit implies that the 
$c^{(g_1,g_2)}_i({u})$ are finite degree polynomials. 
We will explain the gap condition in more details in the 
context of topological string theory on Calabi-Yau manifolds 
in section \ref{leadingcoefficients}.             

To determine now the holomorphic ambiguity for the pure $SU(2)$ theory, 
we expand the topological amplitudes around the monopole 
point $u=1$. This can be achieved by a S-duality
transformation. Under a S-duality transformation $\tau\rightarrow
-\frac{1}{\tau}$, the shifted $E_2$ transforms with weight 2, and the 
Theta functions transform as  $\theta_2^4\rightarrow -\theta_4^4$,
$\theta_3^4\rightarrow -\theta_3^4$, $\theta_4^4\rightarrow -\theta_2^4$. 
The parameter $u$ and $a$ become \footnote{Here we normalize $a_D$ by a 
factor of $2i$ for the consistence of  conventions.}
\begin{eqnarray}
a_D = \frac{2}{3\theta_4^2(\tau)}(E_2(\tau)-\theta_3^4(\tau)-\theta_2^4(\tau)), ~~~
u_D = \frac{\theta_3^4(\tau) +\theta_2^4(\tau) }{\theta_4^4(\tau)}
\end{eqnarray}
We find that in the cusp limit $\tau\rightarrow i\infty$, the parameters go to 
$a_D\sim q^{\frac{1}{2}}\rightarrow 0$, $u_D\rightarrow 1$. This is similar to
the conifold point in the moduli space Calabi-Yau manifolds. We find 
the gap condition \cite{HK1} around this point  completely fixes the 
holomorphic ambiguity.

We obtain compact formulae for higher genus  $F^{(g_1,g_2)}$ similar to those in \cite{HK1} for $F^{(0,g)}$. The genus two formulae are
\begin{eqnarray}
F^{(0,2)} &=& \frac{200X^3-360uX^2+(60u^2+180)X-19u^3-45u}{12960(u^2-1)^2}  \nonumber \\
 F^{(1,1)} &=& \frac{20uX^2-(40u^2+60)X+3u^3+45u}{2160(u^2-1)^2} \nonumber \\
 F^{(2,0)} &=& \frac{10u^2X+u^3-75u}{4320(u^2-1)^2} \label{genus2(2.7)}
\end{eqnarray}
We note that $X,u$ are modular invariant
under the monodromy group $\Gamma(2)\subset SL(2,Z)$ if we shift the 
second Eisenstein series by an anholomorphic piece $E_2\rightarrow \hat E_2=E_2+\frac{6i}{\pi(\bar{\tau}-\tau)}$.

We expand the genus one and genus two formula (\ref{genus1(2.6)}), (\ref{genus2(2.7)}) around the conifold point. 
\begin{eqnarray} 
F_D^{(0,1)} &=& -\frac{1}{12}\log(a_D) +c_{0,1}-\frac{a_D}{2^5}+\mathcal{O}(a_D^2) \nonumber \\
F_D^{(1,0)} &=& \frac{1}{24}\log(a_D) +c_{1,0}-\frac{3a_D}{2^{6}}+\mathcal{O}(a_D^2) \nonumber \\
F_D^{(0,2)} &=& -\frac{1}{240a_D^2} -\frac{a_D}{2^{13}}+\mathcal{O}(a_D^2) \nonumber \\
F_D^{(1,1)} &=& \frac{7}{1440a_D^2} +\frac{3}{2^{11}}+\frac{25a_D}{2^{14}}+\mathcal{O}(a_D^2) \nonumber \\
F_D^{(2,0)} &=& -\frac{7}{5760 a_D^2} +\frac{9}{2^{13}}+\frac{135a_D}{2^{16}}+\mathcal{O}(a_D^2) \label{genus2conifold}
\end{eqnarray}
Here $c_{0,1}$ and $c_{1,0}$ are two irrelevant constants. We see that the
genus two functions satisfy the gap condition with the absence of
$\frac{1}{a_D}$ term. We present gap structure and results 
for $g_1+g_2=3$ in the Appendix \ref{su(2)3}. Our exact formulae (\ref{genus2(2.7)}) sum up the genus two parts of all instanton contributions of the Nekrasov's function. We can check the agreements with Nekrasov's function up to some instanton number, by expanding the expressions around the large complex structure parameter point $u\sim \infty$.

\section{The refined topological string theory}
\label{refinedtopstring}
In this section we discuss general aspects of refined topological 
string theory, such as the description of the expansions in 
terms of refined BPS states and their invariance under complex 
structure deformations. As examples we treat the conifold 
and K3 fibrations. 

First we interpret the general $-\frac{\epsilon_1}{\epsilon_2}=\beta\neq 1$ 
deformation for the BPS states related to topological string theory 
from the generalized Schwinger-Loop amplitude.

\subsection{The Schwinger-Loop amplitude}
\label{BPSamplitude}
 
It will be convenient to define 
\begin{equation} 
\epsilon_{R/L}=\epsilon_\pm=\frac{1}{2} 
(\epsilon_1\pm \epsilon_2)\ . 
\end{equation} 

In~\cite{Morales:1996bp}\cite{IKV}\cite{Antoniadis:2010iq}, it was suggested
to integrate out BPS states in the Schwinger loop amplitude leading to an
$F$-term in $N=2$ supergravity  
\begin{equation} 
R_-^2 T_-^{2m-2} F_+^{2n-2}\ .
\end{equation} 
Here $R_-$ and $T_-$ are the anti-selfdual curvature and anti-self-dual 
graviphoton field strength and  $F_+$ is a self-dual field strength. More
precisely~\cite{IKV} considers for $F_+$ the selfdual part of 
the graviphoton. In this case the amplitude cannot lead to an $F$-term.
In~\cite{Antoniadis:2010iq} for $F_+$ the seldual part of the field 
strength associated to the heterotic dilaton and claim that this 
gives rise to an $F$-term.

The term can be calculated in a 5d M-theory compactification on 
$S^1\times M$ or on an Type II compactification on the Calabi-Yau $M$. 
Following the former picture and denoting the general field strength 
$G=\epsilon_1 dx^1\wedge dx^2+\epsilon_2 dx^3 \wedge dx^4$ then integrating 
out a massive particle of mass $m$ in the representation $\mathcal{R}$ 
of the little group of the 5D Lorentz $SO(4)\sim SU(2)_L \times SU(2)_R$ 
gives the following contribution to the Schwinger-loop amplitude 
\begin{equation}
F(\epsilon_1,\epsilon_2)=-\int_\epsilon^{\infty} \frac{ds}{s}\frac{\textrm{Tr}
  _\mathcal{R} (-1)^{\sigma_L+\sigma_R} e
^{-s m}
e^{-2 i s(\sigma_L\epsilon_{L}+\sigma_R\epsilon_{R})}}{4\left(\sin^2\left(\frac{s\epsilon_L}{2}\right) 
- \sin^2\left(\frac{s \epsilon_R}{2}\right)\right)}\ ,
\label{gapschwingerloop} 
\end{equation}
where we denoted by $\epsilon_{R/L}=\epsilon_\pm=i e G_\pm $ the self-dual or
anti-self-dual part of field strengths coupling to the BPS state 
respectively.

At large complex structure we expect to be able  
to count BPS numbers for the D-brane charges 
$\vec Q=(Q^6,Q^4,Q^2,Q^0)=(1,0,\beta,n)$ with 
$\beta\in H_2(M,\mathbb{Z})$ and $n\in \mathbb{Z}$. 
More precisely in M-theory compactifications on Calabi-Yau 
threefolds $M$ the BPS invariants related to topological string theory 
have been interpreted  as an index in the cohomology of 
the moduli space  $H^*({\cal M}_{\beta})$ of an M2 brane 
wrapping a curve in the class $\beta\in H_2(M,\mathbb{Z})$~\cite{GV}. 
After compactification on the M-theory  $S^1$ the moduli space 
${\cal M}_{\beta}$ can be described equivalently as the one 
of an D2/D0 bound state in the type IIA compactification 
on $M$, where the D2 wraps now the curve and $n$ is the 
degeneracy of the D0 branes.

The $SU(2)_{L/R}$ factors of the little group 
$SO(4)\sim SU(2)_L \times SU(2)_R$ of the 5D Lorentz 
group  of the M theory compactification on $M$ 
act as two Lefshetz actions on the cohomology of 
the moduli space of the brane system $H^*({\cal M}_{\beta})$ 
and these factors of the spacetime group are the 
same that were used in the localization
procedure in~\cite{Nekrasov}. I.e. $\epsilon_L$ 
and $\epsilon_R$ are identified with the 
eigenvalues of the $j_{L/R}^3$ in the corresponding 
$SU(2)_{L/R}$ which label the integer BPS numbers
\footnote{Below we drop the index $3$ on  $j_{L/R}^3$.} 
$n^\beta_{j_L,j_R}$.

One important point here is that $n^\beta_{j_L,j_R}$ 
is not invariant under complex structure deformations 
but only the index 
\begin{equation}
n^\beta_{j_L}=\sum_{j_R} (-1)^{2 j_R}
(2j_R+1)n^{\beta}_{j_L,j_R}\ .
\label{index}
\end{equation}  
This relies on the fact that only the K\"ahler moduli dependence of 
the F-term  $R_-^2T_-^{2g-2}$ defines the invariant topological 
string amplitude and the anti-self-dual graviphoton field strength 
$T_-$ as well as the anti-self-dual curvature 2-form $R_-$ couple 
only to the left spin, while the right spin content enters the 
calculation of $R_-^2T_-^{2g-2}$  merely with its multiplicity 
weighted with $-1$ for fermions and $1$ for bosons 
leading to (\ref{index}). 

In order to compare with the genus expansion of the topological string  
the the left representations have to be organized into 
\begin{equation} 
I_L^n=[(\half )+2(0)]^{\otimes n},
\label{defIr} 
\end{equation}
i.e. one defines 
\begin{equation} 
\sum n^\beta_{j_L,j_R}(-1)^{2 j_R}(2j_R+1)[j_L]=\sum_g n^{\beta}_{g} I_L^g\ .
\label{contr}
\end{equation} 
There is an amusing fact about the expansion $I^n=\sum_{j} c^{2n}_{ j} [j/2]$: The coefficients $c^n_i\in \mathbb{N}$ are the distributions of random walk in the half plane with reflective boundary conditions after $n$ steps, i.e. $c^0_i=\delta_{i,0}$, $c_i^k=0$ for all $k$ and $i<0$ 
$$c^k_i=\left\{\begin{array} {c} c_i^{k-1}+c_{i+1}^{k-1},\quad k \ {\rm even}\\    c_{i-1}^{k-1}+c_{i}^{k-1},\quad k \ {\rm odd}
\end{array}\right. \ .$$
Since $c^{2n}_n=1$ the $[j/2]$ basis can be expressed in terms 
of the $I^r$ with integer coefficient.

Using  (\ref{index}) in (\ref{gapschwingerloop}), 
\begin{equation} 
{\rm Tr}_{I_L^n} (-1)^{\sigma_L} e^{- 2\pi i \sigma_L s}=( 2 \sin(s/2))^{2n}\ ,
\label{trace} 
\end{equation} 
the formula for  mass of the $D2/D0$ brane system $m^2=t+2 \pi i n$ as
well as sum over the $D0$ brane momenta $n$ on the $M$-theory $S^1$ 
yields the formal expression~\cite{GV}
\begin{equation}
\cF^{hol}(\lambda=\epsilon_-,t)=\sum_{g=0}^\infty \sum_{\beta\in H_2(M,\ZZ)}\sum_{m=1}^\infty n^{\beta}_g {1\over m}
\left(2 \sin {m \lambda \over 2}\right)^{2 g-2}  e^{m (\beta, t)}\ .
\label{resschwingerloope-}
\end{equation}
Similarly if one calculates along the same lines  
the $R_-^2 T_-^{2m-2} F_+^{2n-2}$ amplitude one obtains
\begin{equation}
\cF^{hol}(\epsilon_{R/L}, t)=\sum_{{j_L,j_R=0}\atop {m=1}}^\infty \sum_{\beta\in H_2(M,\ZZ)} {n^\beta_{j_L,j_R} \over m} \frac{(-1)^{2 j_L+ 2j_R} 
\left(\sum_{n=-j_L}^{j_L} y_L^{ m n }\right) \left(\sum_{n=-j_R}^{j_R}  y_R^{ m n}\right)
e^{m (\beta, t)}}{4 \left(\sin^2\left(\frac{m \epsilon_L}{2}\right) -
  \sin^2\left(\frac{m \epsilon_R}{2}\right)\right)} \ 
\label{schwingerloope-e+}
\end{equation}
with $y_{L/R}=e^{i \epsilon_{L/R}}$. 
It is  convenient to rewrite (\ref{schwingerloope-e+}) in terms of the
$I_L^{g_L}$ and  $I_R^{g_R}$ basis using 
$\sum {n^\beta}_{j_L,j_R} [J_L,J_R]=\sum \tilde n^\beta_{g_L,g_R} I_L^{g_L}\otimes I_R^{g_R}$
and (\ref{trace}) for the left and the right spin 
\begin{equation}
\begin{array}{rl}
\cF^{hol}(\epsilon_{1/2},t)&=\ds{\sum_{{g_L,g_R=0}\atop {m=1}}^\infty \sum_{\beta\in H_2(M,\ZZ)} {{\tilde n}^\beta_{g_L,g_R} \over m}  
\frac{\sin\left(\frac{m(\epsilon_1-\epsilon_2)}{4}\right)^{2 g_L} 
\sin\left(\frac{m(\epsilon_1+\epsilon_2)}{4}\right)^{2 g_R} 
e^{m (\beta, t)}}
{4 \left(\sin\left( \frac{m \epsilon_1}{2}\right)
(\sin\left( \frac{m \epsilon_2}{2}\right)
\right)}}\\ [3 mm]
&=\ds{\sum_{{g_1,g_1=0}\atop {g_1+g_1=0\ {\rm mod}\ 2}}^\infty  \epsilon_1^{g_1-1}\epsilon_2^{g_2-1} \tilde F_{g_1,g_2} (t)}\ .
\end{array}
\label{schwingerloope1e2}
\end{equation}
Here the ${\tilde F}_{g_1,g_2}$ are easily extracted since at every power of
$\epsilon_{1/2}$ they involve only finitely many $\tilde n^\beta_{g_1,g_2}$. 
We list the first few
\begin{equation} 
\begin{array}{rl}
\tilde F_{0,0}=&\ds{\sum_{\beta} \tilde n^\beta_{0,0} {\rm Li}_3\left(e^{(t,\beta)}\right)}\\
[3 mm]
\tilde F_{0,2}=&\ds{\sum_{\beta} \left(\frac{\tilde n^\beta_{0,1}}{24}+\frac{1}{4}({\tilde n}^\beta_{01} +{\tilde n}^\beta_{1,0})\right) {\rm Li}_1\left(e^{(t,\beta)}\right)}\\[3 mm]
\tilde F_{1,3}=&\ds{\sum_{\beta} \left(\frac{1}{4}({\tilde n}^\beta_{02} -{\tilde n}^\beta_{2,0})\right) {\rm Li}_{-1}\left(e^{(t,\beta)}\right)}\\[3 mm]
\tilde F_{2,2}=&\ds{\sum_{\beta} \left( \frac{\tilde n^\beta_{0,1}}{576}-\frac{1}{96}({\tilde n}^\beta_{02} -{\tilde n}^\beta_{2,0})+ \frac{3}{8}({\tilde n}^\beta_{02} -{\tilde n}^\beta_{2,0})+\frac{1}{8}{\tilde n}^\beta_{1,1} \right) {\rm Li}_{-1}\left(e^{(t,\beta)}\right)}\\
& etc
\end{array}
\end{equation}
and note that generally the Polylogarithm ${\rm
  Li}_{3-g_1-g_2}(x):=\sum_{k=1}^\infty \frac{x^k}{(3-g_1-g_2)^k}$
describes the multi covering of the curve in the class 
$\beta$ contributing to $\tilde F_{g_1,g_2}(t)$. 

Moreover note, that if (\ref{gapschwingerloop}) is the correct starting point 
for the general generalized topological string instanton expansion then there will be no odd powers in 
$\epsilon_1,\epsilon_2$. By comparison with expansion of 
the type (\ref{mixedexpansion}) we see that there are no 
contributions from the instantons to $F^{(n/2,m)}(t)$ for n 
odd. Since  $F^{(1/2,0)}(t)$ is related to the Griffith infinitesimal
invariant in the holomorphic anomaly of~\cite{Krefl:2010fm}, it seems that for the topological 
string the  version of the generalized  holomorphic anomaly 
equation (\ref{generalizedBCOV}) is generally applicable 
for the topological string. 

By geometrical engineering the $SU(2)$ SYM theory with $N_f=1$, for which a non-trivial modification 
of (\ref{generalizedBCOV}) seems necessary~\cite{Krefl:2010fm}, 
is related to the field theory limit of topological string on the 
blow up of $\mathbb{F}_1$~\cite{Katz:1996fh}. One would expect 
to see the non-trivial  contribution by a subtle effect 
in this field theory limit.

\subsection{The gap condition at the conifold point}   
\label{leadingcoefficients}

Here we provide a general derivation of the singular terms in 
the dual expansion near the generic conifold, such as (\ref{genus2conifold}, 
\ref{genus3conifold}). We will be able to explain the gap 
condition as well as computing the leading coefficients. 
Our argument is a generalization of that of~\cite{HKQ}, and has been 
also presented recently in~\cite{Krefl:2010fm}. Basically, 
the singular terms in dual expansion come from integrating out nearly 
massless particles near the conifold point. Generically the
massless BPS state has the charge $\vec Q=(1,0,0,0)$ and has identified 
as a massless extremal black hole~\cite{Strominger}.
In Type II string theory on a Calabi-Yau space it comes from a D3-brane 
wrapping the $S^3$, which shrinks at the conifold 
and its mass squared is $t_c=\int_{S^3} \Omega/t_0$. 
Here $\Omega$ is the holomorphic $(3,0)$ form and $t_0$ is a 
period which starts with the constant one at the conifold. 
In the non-compact limit leading to the Seiberg-Witten gauge theory 
the local reduction of that period becomes $a_D=\int_{S^1} \lambda$, 
where $\lambda$ is the meromorphic Seiberg-Witten differential. 
In the gauge theory the vanishing mass squared is that of 
a magnetic monopole. Following the arguments of Gopakumar and Vafa~\cite{GV} 
and integrating out the nearly massless particle generates the 
singular terms in the dual expansion of 
\begin{equation} 
\label{dualcoefficients}
F(\epsilon_1,\epsilon_2,a_D)=-\int_0^{\infty} \frac{ds}{s}\frac{\exp(-s a_D)}{4\sin(s\epsilon_1/2)\sin(s\epsilon_2/2)} +\mathcal{O} (a_D^0)
\end{equation}
Since the calculation is local we present it only for the 
gauge theory case. For the string case  $a_D$ is simply 
to be replaced with the flat coordinate $t_c$.  

It is straightforward to expand the integrand in small $\epsilon_1$,
$\epsilon_2$ and perform the integral. We compute the first 
few orders \footnote{The logarithmic term $\log(a_D)$ comes from the
 regularization near $s=0$ of the integral  $\int_0^\infty \frac{ds}{s} e^{-sa_D}= -\log(a_D)+\mathcal{O}(a_D^0)$.}
\begin{eqnarray}
F(\epsilon_1,\epsilon_2,a_D) &=& \big{[}-\frac{1}{12}+\frac{1}{24} 
(\epsilon_1+\epsilon_2)^2(\epsilon_1\epsilon_2)^{-1}\big{]}\log(a_D) \nonumber
\\ && 
+ \big{[} -\frac{1}{240}(\epsilon_1\epsilon_2)+\frac{7}{1440}
(\epsilon_1+\epsilon_2)^2-
\frac{7}{5760} (\epsilon_1+\epsilon_2)^4(\epsilon_1\epsilon_2)^{-1} \big{]} \frac{1}{a_D^2} \nonumber \\ && 
+ \big{[} \frac{1}{1008}(\epsilon_1\epsilon_2)^2-\frac{41}{20160}
(\epsilon_1+\epsilon_2)^2(\epsilon_1\epsilon_2) 
+\frac{31}{26880} (\epsilon_1+\epsilon_2)^4  \nonumber \\ &&  -
\frac{31}{161280} (\epsilon_1+\epsilon_2)^6 (\epsilon_1\epsilon_2)^{-1}
\big{]} \frac{1}{a_D^4} +\mathcal{O} (\frac{1}{a_D^6})   +\mathcal{O} (a_D^0) 
\end{eqnarray}
We see the gap structure in the dual expansion around the conifold point, and
the leading coefficients exactly match those in (\ref{genus2conifold},
\ref{genus3conifold}). This universal behavior will enable us 
to fixed the holomorphic ambiguity in the refined topological 
string theory.

\subsection{$B$-model for the resolved conifold}
\label{resconifold}
The resolved conifold can be represented as line bundle over a 
sphere $\mathcal{O}(-1)\oplus \mathcal{O}(-1)\rightarrow \mathbb{P}^1$.
This is one of simplest local Calabi-Yau models where the Gopakumar-Vafa 
correspondence between topological strings and Chern-Simons gauge theory was
first discovered \cite{GV1998}.  The topological A-model on resolved conifold
is particularly simple and the Chern-Simons gauge theory become a matrix model 
in the small K\"ahler parameter limit. From the B-model perspective, a 
Picard-Fuchs differential equation for the model was provided in \cite{FJ}, 
where the complex structure parameter of the mirror curve is simply 
related to the exponential of the K\"ahler parameter $T$ in the A-model by
$Q=e^{-T}$. One can see the Christoffel symbol of the moduli space metric and the propagator defined in \cite{BCOV} are
rational functions of Q.  Therefore in this case we do not need to  integrate 
the holomorphic anomaly equation in B-model because the higher genus
amplitudes are simply rational functions Q. We can determine 
this rational function by the gap condition near the small K\"ahler parameter
limit $T\sim 0$.

It turns out these ideas are also valid in the refined case. In this case
the geometry supports only the rigid $\mathbb{P}^1$ as smooth curve. As 
it is rigid $[j_R]=[0]$ and as it is genus zero $[j_L]=[0]$. Hence
$n^{\mathbb{P}^1}_{0,0}=1$ and all  other $n^{\mathbb{P}^1}_{j_L,j_R}$
vanish. The specialization of (\ref{schwingerloope-e+}) yields
\begin{eqnarray} \label{conifold3.73}
F=-\sum_{n=1}^{\infty}
\frac{Q^n}{n(q^{\frac{n}{2}}-q^{-\frac{n}{2}})(t^{\frac{n}{2}}-t^{-\frac{n}{2}})} 
\end{eqnarray}
Here $Q=e^{-T}$ and $q=e^{\epsilon_1}, t=e^{-\epsilon_2}$.  We can easily extract the refined topological string amplitudes as 
\begin{equation}
\log(Z)= \sum_{i,j=0}^{\infty} (\epsilon_1+\epsilon_2)^i(\epsilon_1\epsilon_2)^{j-1}F^{(\frac{i}{2},j)}(Q)
\end{equation}
One can find 
\begin{eqnarray} \label{conifold3.75}
F^{(g_1,g_2)}\sim \sum_{n>0} n^{2g_1+2g_2-3} Q^n={\rm Li}_{3-2g_1-2g_2}(Q) 
\end{eqnarray}
Here for convenience we only consider the instanton part of the amplitudes, and the classical contribution and constant map contributions at $g>2$ can be easily accounted for. It is possible to sum the infinite series and we found at genus one 
\begin{eqnarray}
F^{(0,1)}=-\frac{1}{12}\log(1-Q),~~~ F^{(1,0)}=  \frac{1}{24}\log(1-Q)
\end{eqnarray}
From the B-model perspective we can compute the higher genus $F^{(g_1,g_2)}$ by requiring it to be a rational function of the form 
\begin{eqnarray}
F^{(g_1,g_2)} (Q)= \frac{\sum_{n=1}^{2g_1+2g_2-3} c_n Q^n}{(1-Q)^{2g_1+2g_2-2}}
\end{eqnarray}
where we have used the boundary condition $F^{(g_1,g_2)} (Q)\sim \frac{1}{T^{2g_1+2g_2-2}}$ when $T\sim 0$ and $Q=e^{-T}\sim 1$, and $F^{(g_1,g_2)} (Q)$ vanishes in both limits $Q\sim 0$ and $Q\sim +\infty$. Furthermore, the following gap condition can completely fix the polynomial in the numerator of $F^{(g_1,g_2)} (Q)$
\begin{eqnarray}
F^{(g_1,g_2)} (Q) \sim  \frac{1}{T^{2g_1+2g_2-2}}+\mathcal{O}(T^0)
\end{eqnarray}
where the leading coefficients can be found either from the expansion in
(\ref{conifold3.73}) 
or the analysis from integrating massless charged particles in 
section \ref{leadingcoefficients}. We find for example 
\begin{eqnarray}
&& F^{(0,2)} =- \frac{Q}{240(1-Q)^2}, ~ F^{(1,1)} =\frac{7Q}{1440(1-Q)^2}, ~ F^{(2,0)} =- \frac{7Q}{5760(1-Q)^2} \nonumber \\
&& F^{(0,3)}= \frac{Q(1+4Q+Q^2)}{6048(1-Q)^4}, ~~~  F^{(1,2)}= -\frac{41Q(1+4Q+Q^2)}{120960(1-Q)^4}, \nonumber \\
&& F^{(2,1)}= \frac{31Q(1+4Q+Q^2)}{161280(1-Q)^4}, ~~~  F^{(3,0)}= -\frac{31Q(1+4Q+Q^2)}{967680(1-Q)^4} 
\end{eqnarray}
In this case the gap condition is understood as coming from the matrix model
description at the small K\"ahler parameter 
limit. On can also see from (\ref{conifold3.75}) that the higher 
genus amplitudes can be directly obtained from genus one 
amplitude by operating with the operator $\Theta^{2g-2}$, 
where $\Theta=Q\frac{d}{dQ}=-\frac{d}{dT}$.  The operator 
$\Theta^{2g-2}$ transform the logarithmic singularity 
$\log(T)$ at genus one to $\frac{1}{T^{2g-2}}$ at 
genus $g$, and also determine the rational function form 
of the higher genus amplitudes.

\subsection{The index and complex structure deformations} 
Let us point out how complex structure specialization 
can lead to different models
for ${\cal M}_\beta$ for which the individual 
$n^{\beta}_{j_L,j_R}$ change, but not the index.
Particular simple examples occur for rational 
curves embedded with degree one~\cite{Katz:1996fh}\cite{Katz}:  
Calabi-Yau hypersurfaces $M$ in weighted projective spaces with a 
$Z_2$ singularity over a smooth genus $g$ curve ${\cal C}_g$, 
e.g. the octic in $\mathbb{P}(1,1,2,2,2)$ where ${\cal C}_3$ is the 
degree 4 hypersurface depending on the last three 
coordinates, contain after resolving the $\mathbb{Z}_2$ 
singularity by an $\mathbb{P}^1$ a rational fibration 
over ${\cal C}_g$. We want to discuss the moduli space 
and the associated BPS numbers of the smooth rational curve 
in the fibration. It represents the basis $[B]$ of the 
hypersurface $M$ viewed as an $K3$ fibration. Because of the 
rational fibration the moduli space of the fiber 
$\mathbb{P}^1$ is ${\cal M}_{[B]}={\cal C}_g$, with 
Dolbeault homology dimensions $\begin{array}{ccc}&1& \\ g&&g \\ &1& \end{array}$.
The $\mathbb{P}^1$ represents the highest 
(and lowest) left spin for a curve in the class $[B]$. 
It  is $[0]_L$ in this case. Since the 
right Lefshetz action  $SU(2)_R$ for 
the highest left spin is just the usual 
Lefshetz action on the deformation space ${\cal C}_g$
~\cite{GV}\cite{Katz:1996fh}, one can read off 
immediately the right representation as 
\begin{equation} 
\left[\frac{1}{2}\right]_R+2 g [0]_R\ .
\end{equation}

One can then show that ${\cal C}_g$ exists only for the toric 
embedding of the hypersurface, which freezes $g-1$ 
complex structure moduli to fixed values. If one 
considers general complex structure 
deformations, the so called non-toric deformations, 
a superpotential of degree $2g-2$ develops~\cite{Kachru:2000ih}, which restricts 
the $\mathbb{P}^1$ to sit at $2g-2$ points hence 
the right spin content is now the one for 
${\cal M}_\beta=(2g-2)$ points, i.e. 
\begin{equation}
(2 g-2) [0]_R \ .
\end{equation} 
The weighted sum yields the Euler number of the deformation space 
with a sign, i.e.  $n^\beta_{J^{\rm max}_L}$ equals 
$n^\beta_{g^{\rm max}}=(-1)^{{\rm dim}({\cal M}_{\beta})} e( {\cal M}_{\beta} )$ and yields in the cases discussed above for which 
$g^{\rm max}=0$ invariantly $n_{0}^{[B]}=2g-2$. Related 
considerations for rational curves on the quintic~\cite{Katz} 
show generally that at complex 
codimension one loci in the complex moduli space the moduli space of the 
rational curves embedded with degree one can jump 
from isolated points to higher genus curves in a way 
which preserves the index. Generally the 
virtual dimension of the moduli of 
holomorphic curves is zero, however even for the most 
general complex structure deformation the 
actual dimension of the moduli space can be 
positive.      

The calculation of $n^\beta_{J_L,J_R}$ is a well defined  
but difficult problem on compact Calabi-Yau spaces, which 
sheds light e.g. on the deformation space of 
holomorphic curves. However for compact Calabi-Yau 
spaces there can be in general no generating function depending just 
on the K\"ahler moduli. To avoid this problem one 
can try fix the complex moduli in a
canonical way.  The most obvious possibility is to 
consider local Calabi-Yau, which have no complex 
structure moduli. Another canonical choice can 
arise for decompactification limit of regular $K3$ 
fibrations.

\subsection{Refined BPS state counting on K3 fibrations} 
 
Here we  discussed a refined G\"ottsche formula, which 
incorporates the left and the right spin degeneracies 
and relate them to an one loop amplitude in 
topological string theory.     

\subsubsection{The refined G\"ottsche Formula}  
\label{refinedgoettsche}
Geometrical Lefshetz decompositions yielding 
only the index have been defined in models 
for ${\cal M}_{\beta}$  and checked in~\cite{Katz:1996fh}
using the Abel-Jacobi map for a variety of 
geometric settings.  Specially simple 
situation arise for curves in surfaces in a CY threefold. 
The easiest example is $K3\times T_2$ where one specializes 
to classes in the $K3$. Strictly speaking this case is 
degenerate, because as far as Gromov-Witten - and Gopakumar-Vafa invariants  
are concerned, there is a multiplicative zero coming 
from the $T_2$. However the description of the moduli  
space of BPS states below finds application for CY, which 
are regular $K3$ fibrations. And in this case a relation 
to  Gromov-Witten invariants exists.  The moduli space for 
the BPS states  of $(D2,D0)$  brane system  
with charge $(\beta,g)$ is the canonical resolution $S^{[g]}$ 
of the Hilbert scheme  of $g$ points, i.e. $S^{\otimes g}$ 
divided by the permutation group ${\rm Sym}^{g}$. The dependence of 
the BPS invariant on the class $\beta$ is only via $\beta\cdot \beta=2g-2$ 
and the $g$ points correspond to the nodes of the general genus $g$  
curve and can be interpreted as positions of the D0 branes.  

For general $S^{[g]}$  G\"ottsche derived a generating 
function $P(X,z)=\sum_i b_i(X) z^i$ capturing the Betti 
numbers of all  $S^{[g]}$ 
\begin{equation} 
\sum_{g=0}^\infty  P(S^{[g]},z)q^g=
\prod_{m=1}^\infty {(1+ z^{2m -1} q^m)^{b_1(S)} (1+ z^{2m + 1}
  q^m)^{b_1(S)}\over (1-z^{2m-2} q^m)^{b_0(S)} (1-z^{2m}
  q^m)^{b_2(S)}(1-z^{2m+2} q^m)^{b_0(S)}}\ . 
\label{goettsche}
\end{equation} 
This can be interpreted as partition function $b_1(S)$ chiral fermions 
and  $b_0(S)+b_2(S)$ chiral bosons, whose oscillators are in addition 
distinguished by the ordinary $SU(2)$ Lefshetz  charge $j_3$. For $z=-1$ that specializes 
to  
\begin{equation} 
\sum_{g=0}^\infty  e(S^{[g]})t^g=\prod_{m=1}^\infty (1-
q^m)^{-e(S)}=\frac{q^\frac{e(S)}{24}}{\eta(q)^{e(S)}}  
\label{euler}
\end{equation} 
and for $K3$, where there is no odd cohomology and $\chi(K3)=24$,  the 
formula can be explained within heterotic type II/duality in six 
dimensions, as counting literally the energy degeneracy of the $24$ 
left (l) moving bosonic oscillators $\alpha_{-k}$ in the index 
${\rm Tr} (-1)^{\bar F_r} q^{L_0-{c\over 24} } q^{\bar L_0-{\bar c\over 24}}$ 
of the heterotic string theory~\cite{Yau:1995mv}.

Let us assume that $b_1(S)=0$ and $b_0(S)=1$ which is true 
for the relevant cases. Then the picture can be refined to implement  
the left and right $SU(2)_L \times SU(2)_R$ quantum numbers on surfaces $S$
by assigning to all bosonic oscillators $\alpha_{-k}$ instead of 
the representation $(b_2(S)+1)[0]+[1]$  in the
diagonal $SU(2)_{L+R}$, which lead to (\ref{goettsche}), the $SU(2)_{L}\times SU(2)_{R}$ 
representation~\cite{Hosono:1999qc}\cite{Katz:1999xq}
\begin{equation} 
\alpha_{-k}: \ \   b_2(S) [0,0]+\left[\frac{1}{2},\frac{1}{2}\right]\ .
\label{reprsu2xsu2}
\end{equation} 
Let us define 
\begin{equation}
G^S(q,z_L,z_R):=\sum_{g=0}^\infty P(S^{[g]},z_L,z_R) q^g
\end{equation} 
with $P(X,z_L,z_R)=\sum_{J_L,J_R} b_{J_L,J_R}(X) z_L^{J_L} z_R^{J_R} $.
The generalization of (\ref{goettsche}) to the representation
(\ref{reprsu2xsu2}) for surfaces with $b_1(S)=0$ reads~\cite{Hosono:1999qc}  
\begin{equation} 
\begin{array}{rl}
\ds{\sum_{g=0}^\infty  P(S^{[g]},z_L,z_R) q^g}=&\displaystyle{ \prod_{m=1}^\infty {1
    \over (1 -(z_L z_R)^{m-1} q^m) (1-(z_L z_R)^{m+1} q^m)(1-(z_L z_R)^{m} q^m)^{b_2(S)-2} }} \\
                                      &\times \displaystyle{{1
                                       \over (1-z_R^2 (z_L z_R)^{m-1}q^m)
(1-z_L^2 (z_L z_R)^{m-1}q^m)}}\ . 
\end{array}
\label{refinedgoettsche}
\end{equation}

From the description of the Lefshetz decomposition of the cohomology of 
the moduli space (\ref{refinedgoettsche}) one can get the genus 
expansion of the topological string in terms of the $n^{\beta}_{g}$ for $K3$ 
can using (\ref{defIr},\ref{trace})~\cite{Hosono:1999qc}\cite{Katz:1999xq} 
\begin{equation} 
G^{K3}(q/y,y,1)=\prod_{n=1}^\infty {1\over { (1-q^n)^{20} (1- y q^n)^2 
(1-y^{-1}q^n)^2}}=\sum_{g=0,\beta}^\infty (-1)^g 
 n^\beta_g (y^{1\over 2} -y^{-{1\over 2}})^{2 g} q^\beta\ . 
\label{BPS}
\end{equation} 

While (\ref{BPS}) counts quantities, which are invariant 
under complex structure deformations, (\ref{refinedgoettsche}) 
contains more complete information of the  $H_({\cal M}_{\beta})$ 
cohomology for a fixed complex structure. The $n^{\beta}_{j_L,j_R}$ 
encode the information of the deformed 
$\Omega$-background 
$\epsilon_1\neq -\epsilon_2$, i.e. $\epsilon_+\neq 0$. 

\subsubsection{Heterotic/Type II duality} 
\label{het/typeII}
Next we compare the result (\ref{refinedgoettsche}) with 
the modified heterotic string one loop contribution
suggested by~\cite{Morales:1996bp}\cite{Antoniadis:2010iq}. 
Let us point out the difference of the latter to the heterotic 
one-loop integral, which leads to the successful evaluation
of BPS invariants in K3-fibered Calabi-Yau 
spaces~\cite{AGNT},\cite{Marino:1998pg}
\cite{Klemm:2004km}\cite{Klemm:2005pd}. Here the integral is 
over the fundamental region of the WS-torus parameterized 
by $\tau=\tau_1+i \tau_2$ 
\begin{equation} 
F_g(t)=\int_{\cal F} \frac{\dd^2 \tau}{\tau_2} \tau^{2 g-2} \sum_{J} {\cal I}^g_J
\label{oneloop1}
\end{equation}   
where 
\begin{equation}
{\cal I}_J^g=\frac{\hat {\cal P}_g}{Y^{g-1}}\bar \Theta^g_J(q) f_J(q)\ .
\end{equation}
The sum over $J$ labels orbifold sectors for which
$\bar \Theta^g_J(q)$ is orbifold projection of the Siegel-Narain $\Theta$-function with $p_r^{2g-2}$ insertions and $f_J$ capture 
the oscillators contributions in the orbifold sectors. This sum
will combine to a modular form of appropriate weight, see 
(\ref{modform}). The amplitude 
depends only on the vector moduli via their 
occurrence in $\Theta^g_J(q)$ and $Y=e^{-K}$, where $K$ is the 
K\"ahlerpotential of the vector moduli metric. 
It is possible to write down all  terms  using\footnote{One could also use (\ref{theta}) to write the total amplitude $F(\lambda,t)=\sum_{\lambda=0^\infty} \lambda^{2g-2} F_g(t)$ directly as integral, but the notation get more clumsy.} 
\begin{equation} 
e^{-\frac{\pi \lambda^2}{\tau_2}}\left(\frac{2 \pi \eta^3 \lambda}{
\theta_1(\lambda|\tau)}\right)^2=\sum_{g=0} (2 \pi \lambda)^{2g} \hat {\cal P}_g=-\exp\left(2 \sum_{k=1}^\infty\frac{\xi(2 k)}{k} \hat E_{2 k}(\tau) \lambda^{2 k}\right) \ . 
\label{theta} 
\end{equation}
Here $\hat E_{2k}$ are the holomorphic Eisenstein series $E_{2k}$ for $k>1$
and $\hat E_2=E_2-\frac{3}{\pi \tau_2}$ the almost holomorphic second
Eisenstein series, i.e. all $\hat E_{2k}$ transform as modular 
forms of weight ${2 k}$. 

In~\cite{Morales:1996bp}\cite{Antoniadis:2010iq} it was suggested to couple
the BPS states in the Schwinger loop amplitude to an additional 
self-dual matter vector field strength $F_+$, i.e. they consider the one loop  
amplitude $R_-^2 T_-^{2g-2} F_+^{2n-2}$. The effect is merely to split
(\ref{theta}) as       
\begin{equation} 
e^{-\frac{\pi (\epsilon^2_-+\epsilon^2_+)}{\tau_2}}
\left(\frac{2 \pi (\epsilon_- +\epsilon_+)\eta^3}{
\theta_1((\epsilon_-+\epsilon_+)|\tau)}\right)\left(\frac{2 \pi (\epsilon_- -\epsilon_+)\eta^3}{
\theta_1((\epsilon_--\epsilon_+)|\tau)}\right)=\sum_{m,n} (2 \pi
(\epsilon_-+\epsilon_+))^{2 m} (2 \pi (\epsilon_--\epsilon_+))^{2 n} 
\hat {\cal P}_{m,n}\ , 
\label{theta2} 
\end{equation}
here 
\begin{equation} 
\hat {\cal P}_{m,n}={\cal S}_m(x_1,\ldots, x_m){\cal S}_n(x_1,\ldots, x_n)\ 
\end{equation}      
are almost modular forms of weight $2m + 2 n$. Concretely  
$x_k=\frac{|B_{2k}|}{2 k (2 k)!} \hat E_{2k}$ and 
${\cal S}_m(\underline{x})$ is defined by  $\exp(\sum_{n=1} x_i
z^{i})=\sum_{n=0}^\infty {\cal S}_m({\underline x}) z^m$.  
This allows to define
\begin{equation} 
F_{m,n}(t)=\int_{\cal F} \frac{\dd^2 \tau}{\tau_2} \tau^{2 (m+n)-2} \sum_{J} {\cal I}^{m,n}_J
\label{oneloop2}
\end{equation}   
with
\begin{equation}
{\cal I}_J^{m,n}=\frac{\hat {\cal P}_{m,n}}{Y^{m+n-1}}\bar \Theta^{m+n}_J(q) f_J(q)\ .
\end{equation}
Now we can point out the difference in the calculation of the heterotic one-loop amplitude (\ref{oneloop1}) and (\ref{oneloop2}).
First for $\epsilon_+=0$ (\ref{oneloop2}) specializes to (\ref{oneloop1}) and this has been calculated for many examples 
of heterotic/Type II pairs starting with the $STU$ model 
in~\cite{Marino:1998pg} in more general 
situations~\cite{Klemm:2004km}\cite{Klemm:2005pd}
\cite{Weiss:2007tk}. If the Calabi-Yau space is a regular 
$K3$-fibration the result for the one-loop amplitude 
in the holomorphic limit is expressed using the expansion\cite{Klemm:2004km}  
\begin{equation} 
\begin{array}{rl} 
\ds{{\cal G}^{hol}_{K3}(\lambda,t)}&=\ds{\frac{M(q)}{q} \left(\frac{\lambda}{2 \sin(\frac{\lambda}{2})}\right)^2 G^{K3}(q/y,y,1)}\\ 
                                  &=\ds{\sum_{g=0,d=-1}^\infty c_g(d) \lambda^{2g-2} q^d}\ , 
\label{genoneloop1}
\end{array}
\end{equation}
where  $y=e^{i \lambda}$ and the K\"ahler of K3 enters via $q=e^{2\pi i t}$, by
\begin{equation} 
\cF^{hol}_{K3}(\lambda,t)=
\sum_{g=0}^\infty \sum_{\alpha\in H^{prim}_2(K3,\mathbb{Z})} \lambda^{2g-2} c_g(\alpha^2/(2r)) \frac{Li_{3-2g} (e^{(\alpha, t)} )}{(2\pi i)^{3-2g} }\  . 
\label{FK3}
\end{equation}
The function $\frac{M(q)}{q}$ has been determined in many cases. $r$ depends 
on the Picard-Lattice of the generic $K3$ fiber. First
one notes that there will be always a factor
$\frac{1}{\eta(q)^{24}}=\frac{1}{q \prod_{n=1}^\infty (1-q^n)^{24}}$ in 
${\cal G}^{hol}_{K3}(\lambda,t)$ which comes from the left moving bosonic 
oscillator modes of the heterotic string\footnote{In fact applied to the 
six-dimensional heterotic on $T^4$ versus type IIA on $K3$  duality  
it reproduces the famous (\ref{euler}) as counting function of nodal
curves on $K3$ as observed by  Yau and Zaslow.}.  
It is therefore convenient to define
\begin{equation}
\frac{M(q)}{q}=\frac{\Theta(q)}{q \prod_{n=1}^\infty (1-q^n)^4}\ .
\end{equation}          
$\Theta(q)$ is a form under, in general, a subgroup $SL(2,\mathbb{Z})$ 
of weight $11-\frac{r}{2}$ where $r$ is the rank of the Picard 
Lattice of the $K3$. E.g. for the $ST$ $(r=2)$ and $STU$ $(r=1)$  
model one has 
\begin{equation} 
\Theta^{ST}(q)=\theta_3(\tau/2) E_4 F_6, \qquad    \Theta^{STU}(q)=E_4 E_6 \ 
\label{modform}
\end{equation} 
and $F_6=E_6-2 F_2(\theta^4_3(\tau/2)-2 F_2)(\theta^4_3(\tau/2)-16 F_2)$, 
where $F_2(q)=\sum_{n\in \mathbb{Z}_+,odd} \sigma_1(n)
q^{\frac{n^2}{4}}$. Much more general examples have been 
discussed in~\cite{Klemm:2004km}\cite{Haghighat:2009nr}.

It is worthwhile to stress that the $c_g(\alpha^2/2)\in \mathbb{Z}$ 
are not the BPS invariants $n_g^\alpha$. To get the latter we have to 
compare (\ref{FK3})  with (\ref{resschwingerloope-}) for 
classes  $\beta\in H_2(K3,\ZZ)$.

Going over the calculation leading to (\ref{oneloop1}) 
one recognized the difference for the evaluation of 
(\ref{oneloop2}) does not affect the $\frac{\Theta(q)}{q}$ 
part, which is clear as it is completely determined by the 
genus zero contributions, which are deformation invariant. 
Incorporating the $\epsilon_+$ deformation we can use  
as before the Jacobi triple function identity. 
\begin{equation}
\theta_1(z,\tau)=-2 q^\frac{1}{8} \sin(\pi z) \prod_{m=1} (1-q^m) (1-2 \cos( 2\pi z) q^m+ q^{2m})
\end{equation}
to obtain
\begin{equation} 
\begin{array}{rl} 
\ds{{\cal G}^{hol}_{K3}(\epsilon_{R/L},t)}=&\ds{\frac{M(q)}{q}
  \frac{1}{4 \left(\sin^2\left(\frac{\epsilon_L}{2}\right) - \sin^2\left(\frac{\epsilon_R}{2}\right)\right)}\prod_{n> 0} 
\frac{1}{(1-y_L y_R q^n) (1-y^{-1}_L y^{-1}_R q^n)}} \\ & \ds{\frac{1}{(1-y_L
  y^{-1}_R q^n) (1-y^{-1}_L y_R q^n) (1-q^n)^{20}   }}\\ [6 mm]
                                 =& \ds{\frac{M(q)}{q}
 \frac{1}{4 \left(\sin^2\left(\frac{\epsilon_L}{2}\right) - \sin^2\left(\frac{\epsilon_R}{2}\right)\right)} G^{K3}(q/(y_L y_R),y_L,y_R)}\\ 
                                 =&:\ds{\sum_{n,m=0,d=-1}^\infty c_{n,m} (d)
                                   \epsilon_1^{m-1}\epsilon_1^{n-1}  q^d}\ . 
\label{genoneloop2}
\end{array}
\end{equation}
Here we rescaled $\epsilon_{R/L}$ by $2 \pi i$ and set
$y_{R/L}=e^{i\epsilon_{R/L}}$. The formula gives the 
desired interpretation of the BPS contributions to the heterotic 
one-loop integral~\cite{Morales:1996bp}\cite{Antoniadis:2010iq} 
in terms of the Lefshetz decomposition the moduli spaces of 
curves on the $K3$ fiber. The free energy is then given by    
\begin{equation} 
\cF^{hol}_{K3}(\epsilon_{L/R},t)=
\sum_{m,n=0}^\infty \sum_{\alpha\in H^{prim}_2(K3,\mathbb{Z})} 
\epsilon_1^{m-1}\epsilon_2^{n-1} c_{m,n}(\alpha^2/(2 r)) \frac{Li_{3-m-n}
  (e^{(\alpha, t)} )}{(2\pi i)^{3-m - n} }\  . 
\label{FK3fine}
\end{equation}
To read of the $n^\alpha_{j_L,j_R}$ one compares (\ref{schwingerloope1e2}) for classes in the $K3$ fiber with  
(\ref{FK3fine}) and then re-express the result in terms 
of  the $(J_L,J_R)$ basis. Let us label the classes in 
the $K3$ fibre of the STU model by $(1,n)$, we get then  
\begin{equation} 
\begin{array}{rl} 
n=1: & 488 {\bf \left(0,0\right)}- 2  
{\bf \left(\frac{1}{2},\frac{1}{2}\right)} \\ [ 3 mm] 
n=2: & 280962 {\bf \left(0,0\right)}+ 486  {\bf \left(\frac{1}{2},\frac{1}{2}\right)} - 2 {\bf \left(1,1\right)}\\ [3 mm]
n=3: & 15298438 {\bf \left(0,0\right)}+ 281448  {\bf \left(\frac{1}{2},\frac{1}{2}\right)} + 486 {\bf \left(1,1\right)}-2 {\bf \left(\frac{3}{2},\frac{3}{2}\right)} -2 ({\bf \left(1,0\right)} +{\bf \left(0,1\right)}) \\ [3 mm]
n=4: &  410133612 {\bf \left(0,0\right)}+   16209886 {\bf \left(\frac{1}{2},\frac{1}{2}\right)} +   281446 {\bf \left(1,1\right)}+   486 {\bf \left(\frac{3}{2},\frac{3}{2}\right)} -2 {\bf \left(2,2\right)} + \\ [ 2mm ]
&  486({\bf \left(1,0\right)} +{\bf \left(0,1\right)}) -2 
\left({\bf \left(\frac{1}{2},\frac{3}{2}\right)} +{\bf \left(\frac{1}{2},\frac{3}{2}\right)}\right) 
\end{array}
\label{resSTU}
\end{equation}

From (\ref{oneloop1}) and (\ref{theta}) one can work out
the full an-holomorphic dependence of (\ref{FK3}), as was 
done for the STU model in Appendix C.1 of~\cite{Grimm:2007tm}, 
and check that it is compatible with the holomorphic anomaly 
equation of~\cite{BCOV}. It is not hard to trace the 
anholomorphic dependence under the factorization of 
(\ref{theta}) into (\ref{theta2}) and show that it 
leads to the sum structure in the second term of the 
right hand side in the generalized holomorphic 
anomaly equation (\ref{generalizedBCOV}).  Results 
for other regular K3 fibrations are obtained similarly.

As it is clear from the explanations in the introduction 
to this chapter, it cannot be true in general that the 
description of right handed BPS states does not depend 
on the complex structure, i.e. the hyper multiplets in 
the heterotic string. One possible interpretation is 
that in the strict weak coupling limit, which corresponds 
to infinite volume of the base $\mathbb{P}^1$ the description and 
the right Lefshetz decomposition of the moduli space of 
curves in the K3 fiber is invariant. This would explain, 
why the authors~\cite{Morales:1996bp}\cite{Antoniadis:2010iq} 
don't find hyper multiplet dependence in their perturbative 
calculation. From the examples of the K3 fibered hypersurfaces 
in $\mathbb{P}^4(1,1,w_1,w_2,w_3)$, discussed in the introduction  
one can conclude that the hyper multiplet must couple in the 
non-perturbative sector of the heterotic string. In the next 
section we will apply the above results to a situation where 
the decoupling of the complex  moduli is obvious in the 
geometrical context.

\section{Local Calabi-Yau manifolds} 
\label{localmodels}

One obvious possibility to decouple the complex moduli is to 
look at local models. In particular for all del Pezzo surfaces 
$B$ embedded in a Calabi-Yau three manifold one can take a local 
limit in which the local non-compact Calabi-Yau is described 
by the total space of the canonical line bundle 
${\cal O}(K_B)\rightarrow B$. We start with the del Pezzo's 
which are elliptic fibrations over $\mathbb{P}^1$  
           
\subsection{A rational elliptic surface: half K3}
\label{halfk3}

In~\cite{KMV} simple expressions for the BPS number 
generating function in terms of $SL(2,\mathbb{Z})$ modular forms 
for  the rational elliptic surface $B_n$ embedded in a 
Calabi-Yau threefold $M$ were found. In the simplest example 
$B_9$ was embedded in an elliptic fibration over the Hirzebruch 
surface $\mathbb{F}_1$ and two of the ten classes in $H_2(B_9,\mathbb{Z})$ 
were independent in $M$, namely the base $P$ and the fiber 
$F$ of the elliptically fibered $B_9$.
 
The expression found in~\cite{KMV} described genus zero Gromov-Witten 
invariants $r^{P+n F}_0$, which are wound one times around the base $P=\mathbb{P}^1$ 
and $n$ times around the elliptic fiber 
\begin{equation}
H^{(0)}_1(q)=\sum_{n=0}^\infty r^{P+n F}_0 q^n
  =\frac{q^\frac{1}{2} E_4(q)}{\eta^{12}(q)} \ 
\end{equation}       
and has given an explanation in terms of tensionless 
strings~\cite{KMV}, see also~\cite{Ganor:1996gu}. Note 
that $r^{P+n F}_0=n^{p+ n F}_0$ by virtue of (\ref{resschwingerloope-}).    
 
Higher genus curves of genus $g$ and degree one in the base have 
a simple geometry. They are copies of the elliptic fiber over 
$g$ points in the basis $P=\mathbb{P}^1$. The moduli space 
${\cal M}_{P+n F}$ consist of the possible position 
of these fibers on the base $\mathbb{P}^1$ and the $U(1)$ 
connection on each of the $g$ elliptic fibers. By T-duality 
on the fiber fixing a $U(1)$ connection equivalently 
corresponds to picking point on the dual $T^2$. Therefore 
one expects that the $SU(1)_L\times SU(1)_R$ decomposition of 
the cohomology  of the moduli space of the $P+nF$ curves in this 
is described again by the generating function~\cite{Hosono:1999qc} 
$G^{B_9}(q/(y_L y_R),y_L,y_R)$, with $\chi(B_9)=12$ and 
$b_2(B_9)=10$. Similar as in~(\ref{genoneloop1}) it 
should be supplemented by the $SL(2,\mathbb{Z})$ modular part from 
genus zero. Therefore the result for the refined topological invariants 
is obtained from  
\begin{equation} 
{\cal G}^{B_9}_{hol}= \frac{q^\frac{1}{2} E_4(q)}{\eta^{4}(q)} 
\frac{1}{4 \left(\sin^2\left(\frac{\epsilon_L}{2}\right) - \sin^2\left(\frac{\epsilon_R}{2}\right)\right)} G^{B_9}(q/(y_L y_R),y_L,y_R) 
\end{equation}
by the same steps that lead to (\ref{resSTU})
\begin{equation} 
\begin{array}{rl} 
g=1: & 248 {\bf \left(0,0\right)}+ {\bf \left(\frac{1}{2},\frac{1}{2}\right)} \\ [ 3 mm] 
g=2: & 4125 {\bf \left(0,0\right)}+ 249  {\bf \left(\frac{1}{2},\frac{1}{2}\right)} +1 {\bf \left(1,1\right)}\\ [3 mm]
n=3: & 35001 {\bf \left(0,0\right)}+ 4374  {\bf \left(\frac{1}{2},\frac{1}{2}\right)} + 249 {\bf \left(1,1\right)}+ {\bf \left(\frac{3}{2},\frac{3}{2}\right)} +({\bf \left(1,0\right)} +{\bf \left(0,1\right)}) \\ [3 mm]
n=4: &  217501 {\bf \left(0,0\right)}+   39375 {\bf \left(\frac{1}{2},\frac{1}{2}\right)} +   4375 {\bf \left(1,1\right)}+   249 {\bf \left(\frac{3}{2},\frac{3}{2}\right)}  +{\bf \left(2,2\right)} + \\ [ 2mm ]
&  249({\bf \left(1,0\right)} +{\bf \left(0,1\right)}) + 
\left({\bf \left(\frac{1}{2},\frac{3}{2}\right)} +{\bf \left(\frac{1}{2},\frac{3}{2}\right)}\right) 
\end{array}
\label{resDP9}
\end{equation}
The new point is that these results can be also interpreted 
as refined gauge theory invariants for N=4 $SU(2)$ theory on 
the half K3~\cite{Minahan:1998vr}. Since $b_+(B_9)=1$ we expect 
the refined individual cohomology numbers of the moduli space of 
gauge theory  instantons to be invariant~\cite{Vafa:1994tf}. 
Moreover one expects by direct integration of a holomorphic 
anomaly in the base degree as in~\cite{Hosono:1999qc} to 
be able to extend this result to higher rank gauge groups 
and to further classes of the $B_n$ surfaces~\cite{Minahan:1998vr}.

\subsection{The refined topological string on local $\mathbb{P}^1\times \mathbb{P}^1$}
\label{P1P1}

The local  $\mathbb{P}^1\times \mathbb{P}^1$ is a well studied non-compact Calabi-Yau manifold. This Calabi-Yau geometrically engineers $SU(2)$ Seiberg-Witten theory, and the topological string partition function reduces to the Nekrasov partition function at 
$\epsilon_1+\epsilon_2=0$ in a certain limit. One can also use the refined topological vertex to construct the \textit{refined} topological string amplitude that reduces to Nekrasov partition function in general $\Omega$ background. This was studied in \cite{IKV, AK}. It is found that the world-sheet instanton contribution to the refined topological partition function for this model is 
\begin{eqnarray} \label{P1P1Z}
Z(Q_1,Q_2,t,q) &=& \sum_{\nu_1,\nu_2}Q_1^{|\nu_1|+|\nu_2|} q^{||\nu_2^t||^2}t^{||\nu_1^t||^2}\tilde{Z}_{\nu_1}(t,q)
\tilde{Z}_{\nu_2^t}(t,q)\tilde{Z}_{\nu_2}(q,t)\tilde{Z}_{\nu_1^t}(q,t) \nonumber \\
&\times & \prod_{i,j=1}^{\infty} (1-Q_2t^{i-1-\nu_{2,j}}q^{j-\nu_{1,i}})^{-1}(1-Q_2q^{i-1-\nu_{1,j}}t^{j-\nu_{2,i}})^{-1}
\end{eqnarray}
Some explanations of the  notations are the followings. The $Q_1=e^{2\pi i T_1}, Q_2=e^{2\pi i T_2}$ are exponentials of the K\"ahler parameters $T_1,T_2$ of the model.  The string coupling constant is refined into two parameters $\epsilon_1, \epsilon_2$ and are related to $q,t$ as $q=e^{\epsilon_1}, t=e^{-\epsilon_2}$. The sum in the above equation (\ref{P1P1Z})  is over all 2D Young tableaux $\nu_1, \nu_2$. A 2D Young tableau is a sequence of non-increasing non-negative integers  $\nu=\{ \nu_{,1}\geq\nu_{,2}\geq \cdots \geq 0\}$, and $\nu^t$ denotes the transpose of the Young tableau $\nu$. Some definitions related to the Young tableaux are the followings 
\begin{eqnarray}
|\nu| &=& \sum_{i=1}^{\infty} \nu_{,i}, \nonumber \\
||\nu||^2  &=& \sum_{i=1}^{\infty} (\nu_{,i})^2,  \nonumber \\
\tilde{Z}_{\nu}(t,q) &=& \prod_{(i,j)\in \nu}(1-t^{(\nu^t)_{,j}-i+1}q^{\nu_{,i}-j})^{-1}
\end{eqnarray} 
We perform the sum over Young tableaux in (\ref{P1P1Z}) to a finite order in
$Q_1, Q_2$. In order to perform the infinite product  exactly, we use the
formula $(1-x)^{-1}=\exp(\sum_{n\geq 0} \frac{x^n}{n})$ to convert the
infinite product to infinite sums of geometric series of $t$ and $q$. To
compare with the B-model calculations from Picard-Fuchs equation and
holomorphic anomaly, we should expand the logarithm of the partition 
function as in (\ref{mixedexpansion}), where $F^{(\frac{i}{2},j)}(Q_1,Q_2)$ depends 
now on the flat K\"ahler coordinates $T_{1,2}$  of the $\mathbb{P}^1\times \mathbb{P}^1$ 
geometry via $Q_{1,2}=e^{2\pi i T_{1,2}}$. As explained at the end of section~\ref{BPSamplitude}, 
only $F^{(i,j)}$ with integers $i,j$ should appear in the expansion. The first
non-trivial contribution from the refinement of the topological 
string is $F^{(1,0)}$. We perform the computation on the 
refined partition function ( \ref{P1P1Z}) to first few orders and the results  at genus one are 
\begin{eqnarray}
F^{(0,1)} &=& -\frac{1}{6}(Q_1+Q_2)-\frac{1}{12}(Q_1^2+4Q_1Q_2+Q_2^2)-\frac{1}{18}(Q_1^3+9Q_1^2Q_2+9Q_1Q_2^2+Q_2^3) \nonumber \\
&& -\frac{1}{24}(Q_1^4+16Q_1^3Q_2-148Q_1^2Q_2^2+16Q_1Q_2^3+Q_2^4)+\mathcal{O}(Q^5) \\
F^{(1,0)} &=& -\frac{1}{6}(Q_1+Q_2)-\frac{1}{12}(Q_1^2+28Q_1Q_2+Q_2^2)-\frac{1}{18} (Q_1^3+153Q_1^2Q_2+153Q_1Q_2^2+Q_2^3) \nonumber \\
&& -\frac{1}{24}(Q_1^4+496Q_1^3Q_2+2204Q_1^2Q_2^2+496Q_1Q_2^3+Q_2^4)+\mathcal{O}(Q^5)  \label{P1P1F10} 
\end{eqnarray}
The genus two results 
\begin{eqnarray}
F^{(0,2)} &=& -\frac{1}{120} \left(Q_1+Q_2\right)-\frac{1}{60} \left(Q_1^2+Q_1 Q_2+Q_2^2\right) 
-\frac{1}{40} \left(Q_1^3+Q_1^2 Q_2+Q_1 Q_2^2+Q_2^3\right)  \nonumber  \\
&& -\frac{1}{30} \left(Q_1^4+ Q_1^3Q_2+5  Q_1^2Q_2^2+ Q_1Q_2^3+Q_2^4\right) +\mathcal{O}(Q^5)  \nonumber \\
F^{(1,1)} &=& -\frac{1}{90} \left(Q_1+Q_2\right)-\frac{1}{90} \left(2 Q_1^2+17 Q_1 Q_2+2 Q_2^2\right) \nonumber \\ && 
-\frac{1}{30} \left(Q_1^3+21 Q_1^2Q_2+21 Q_1 Q_2^2+Q_2^3\right) 
\label{P1P1genustwo} \\ && -\frac{1}{45} \left(2 Q_1^4+77  Q_1^3Q_2-995 Q_1^2 Q_2^2+77  Q_1Q_2^3+2 Q_2^4\right) +\mathcal{O}(Q^5)   \nonumber \\
F^{(2,0)} &=& \frac{1}{360} \left(Q_1+Q_2\right)+\frac{1}{180} \left(Q_1^2-59 Q_1 Q_2+Q_2^2\right)\nonumber \\ &&
+\frac{1}{120} \left(Q_1^3-399 Q_1^2Q_2-399 Q_1Q_2^2 +Q_2^3\right)\nonumber \\ && +\frac{1}{90} \left(Q_1^4-1379  Q_1^3Q_2-7495 Q_1^2 Q_2^2-1379 Q_1Q_2^3+Q_2^4\right)   +\mathcal{O}(Q^5)  \nonumber
 \end{eqnarray}
The $(g_1+g_2)=3$ results are relegated to the Appendix (\ref{p1p13}).

\begin{figure} 
\center
\begin{picture}(0,0)%
\includegraphics{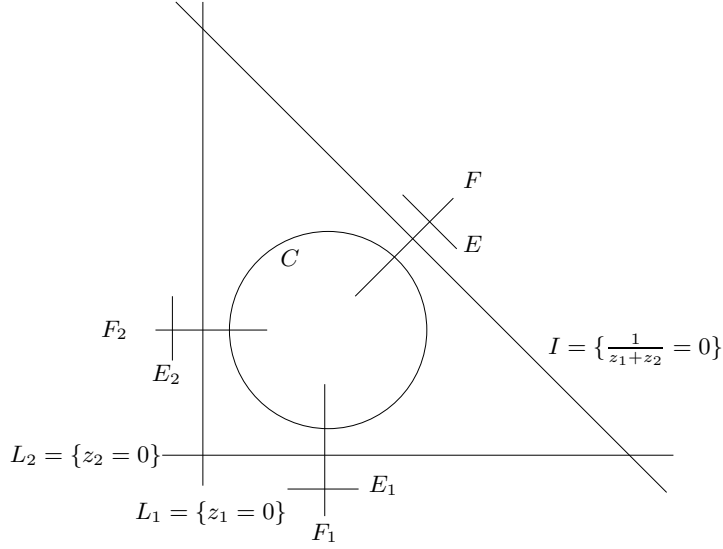}%
\end{picture}%
\setlength{\unitlength}{1865sp}%
\begingroup\makeatletter\ifx\SetFigFont\undefined%
\gdef\SetFigFont#1#2#3#4#5{%
  \reset@font\fontsize{#1}{#2pt}%
  \fontfamily{#3}\fontseries{#4}\fontshape{#5}%
  \selectfont}%
\fi\endgroup%
\begin{picture}(8832,7234)(-764,-7238)
\put(4006,-6541){\makebox(0,0)[lb]{\smash{{\SetFigFont{9}{10.8}{\rmdefault}{\mddefault}{\updefault}$E_1$}}}}
\put(901,-6901){\makebox(0,0)[lb]{\smash{{\SetFigFont{9}{10.8}{\rmdefault}{\mddefault}{\updefault}$L_1=\{z_1=0\}$}}}}
\put(-764,-6136){\makebox(0,0)[lb]{\smash{{\SetFigFont{9}{10.8}{\rmdefault}{\mddefault}{\updefault}$L_2=\{z_2=0\}$}}}}
\put(6391,-4696){\makebox(0,0)[lb]{\smash{{\SetFigFont{9}{10.8}{\rmdefault}{\mddefault}{\updefault}$I=\{\frac{1}{z_1+z_2}=0\}$}}}}
\put(3241,-7171){\makebox(0,0)[lb]{\smash{{\SetFigFont{9}{10.8}{\rmdefault}{\mddefault}{\updefault}$F_1$}}}}
\put(451,-4471){\makebox(0,0)[lb]{\smash{{\SetFigFont{9}{10.8}{\rmdefault}{\mddefault}{\updefault}$F_2$}}}}
\put(5266,-2491){\makebox(0,0)[lb]{\smash{{\SetFigFont{9}{10.8}{\rmdefault}{\mddefault}{\updefault}$F$}}}}
\put(5266,-3346){\makebox(0,0)[lb]{\smash{{\SetFigFont{9}{10.8}{\rmdefault}{\mddefault}{\updefault}$E$}}}}
\put(2836,-3526){\makebox(0,0)[lb]{\smash{{\SetFigFont{9}{10.8}{\rmdefault}{\mddefault}{\updefault}$C$}}}}
\put(1126,-5056){\makebox(0,0)[lb]{\smash{{\SetFigFont{9}{10.8}{\rmdefault}{\mddefault}{\updefault}$E_2$}}}}
\end{picture}%
\caption{Resolved Moduli Space of $\HirzeF_0$}
\label{rms}
\end{figure}

The $F^{(0,g)}$ is the ordinary topological string amplitude and has been well
studied before from B-model using mirror symmetry, see e.g. \cite{HKR}. Here
we include it for the purpose of checking the calculations. The K\"ahler
parameters $Q_1=e^{2\pi i T_1}, Q_2=e^{2\pi i T_2}$ are related to the complex structure parameters $z_1, z_2$ of the mirror manifold through Picard-Fuchs equations. The two Picard-Fuchs operators are 
\begin{eqnarray}
L_1&=& \Theta_1^2 -2z_1(\Theta_1+\Theta_2)(1+2\Theta_1+\Theta_2)  \nonumber \\
L_2 &=& \Theta_2^2 -2z_2(\Theta_1+\Theta_2)(1+2\Theta_1+\Theta_2) 
\end{eqnarray}
where $\Theta_i=z_i\frac{\partial}{\partial z_i},  i=1,2$.  The discriminant
is $z_1z_2\Delta=0$ where the conifold discriminant is given by
\begin{eqnarray}
\Delta=1-8(z_1+z_2)+16(z_1-z_2)^2\ .
\label{conifolddiscp1p1}
\end{eqnarray}
Here we depict the moduli space in Figure \ref{rms} following \cite{HKR}. The conifold loci $C$ is parameterized by $\Delta=0$, and intersect tangentially with the other singular loci $z_1=0$, $z_2=0$, and $\frac{1}{z_1+z_2}=0$. To study the model around the tangent intersection points, we need to blow up the points by adding the extra lines $F, F_1,F_2$ in the Figure \ref{rms}, so that the intersections become normal. In this paper we will study the model around the large volume point $z_1=z_2=0$, and around a generic point on the conifold loci not intersecting with other singular point.

The Picard-Fuchs equations of local models have a constant solution and the K\"ahler moduli $T_1, T_2$ in the mirror are the two logarithmic solutions of the Picard-Fuchs equations around the large volume point $z_1=z_2=0$,
\begin{eqnarray}
2\pi iT_1 &=& \log(z_1) + 2(z_1+z_2) +3(z_1+4z_1z_2+z_2^2)+ \mathcal{O}(z^3) \nonumber \\
2\pi iT_2 &=& \log(z_2) + 2(z_1+z_2) +3(z_1+4z_1z_2+z_2^2)+ \mathcal{O}(z^3)  \label{P1P1mirrormap1}
\end{eqnarray} 
Exponentiate and invert the above series expansion one finds
\begin{eqnarray}
z_1(Q_1,Q_2) &=& Q_1-2Q_1(Q_1+Q_2)+3Q_1(Q_1^2+Q_2^2)+\mathcal{O}(Q^4) \nonumber \\ 
z_2(Q_1,Q_2) &=& Q_2-2Q_2(Q_1+Q_2)+3Q_2(Q_1^2+Q_2^2)+\mathcal{O}(Q^4)  \label{P1P1mirrormap2}
\end{eqnarray}
If the refined topological string amplitudes can be computed from mirror symmetry through the extended holomorphic anomaly equations, we should expect the first non-trivial amplitude from the refinement, $F^{(1,0)}$, to be proportional the logarithm of the discriminant the Calabi-Yau geometry, from our discussion of the pure $SU(2)$ Seiberg-Witten theory. Indeed we check the non-perturbative parts of the logarithm agree with the results  in (\ref{P1P1F10})  from the refined topological vertex calculations
\begin{eqnarray}
\frac{1}{24}\log(\frac{\Delta}{z_1z_2}) &=& -\frac{1}{24}\log(Q_1Q_2) -\frac{1}{6}(Q_1+Q_2)-\frac{1}{12}(Q_1^2+28Q_1Q_2+Q_2^2)  \nonumber \\ && -\frac{1}{18} (Q_1^3+153Q_1^2Q_2+153Q_1Q_2^2+Q_2^3) +\mathcal{O}(Q^4)
\end{eqnarray}

We then compute the higher genus amplitudes in the refined topological string using the extended holomorphic anomaly equations and the boundary conditions at singular points of the moduli space in the mirror. We follow the techniques developed in \cite{HKR} for dealing with multi-parameter Calabi-Yau models. For the local model we consider, the K\"ahler potential is a constant in the holomorphic limit, so the covariant derivative with respect to the vacuum line bundle $\mathcal{L}$  vanishes, and we only need to use the Christoffel connection $\Gamma^i_{jk}$ with respect to the metric in the covariant derivative. In the holomorphic limit, the metric and Christoffel connection can be calculated from the mirror maps (\ref{P1P1mirrormap1}) (\ref{P1P1mirrormap2})
\begin{eqnarray} \label{P1P1Christoffel}
G_{i\bar{j}}\sim \frac{\partial T_j}{\partial z_i}, ~~~  \Gamma_{jk}^i= \frac{\partial z_i}{\partial T_l}\frac{\partial^2 T_l}{\partial z_j\partial z_k} 
\end{eqnarray}

To integrate the holomorphic anomaly equation, we should write the topological string as polynomials of some generators following the approach in \cite{YY, HKQ}. It turns out that for multi-parameter models, it is more convenient to use the propagators $S^{jk}, S^k, S$ as anholomorphic generators of the polynomials \cite{AL}. The propagators were originally introduced by BCOV \cite{BCOV} to integrate the holomorphic anomaly equations, and for local models we only need the two-index propagators, which are defined in terms of the three point coupling as $\bar{\partial}_{\bar{i}}S^{jk}= \bar{C}^{jk}_{\bar{i}}$. The main difference with one-parameter model is that the topological string amplitudes will be polynomials of the anholomorphic generators $S^{jk}$ with the coefficients as rational functions of the moduli $z_i$, where in one-parameter models one can also include a holomorphic generator which is a rational function of the modulus, and the topological string amplitudes would be truly polynomials with constant coefficients. Assuming the anti-holomorphic derivative of the propagators $\bar{\partial}_{\bar{i}}S^{jk}= \bar{C}^{jk}_{\bar{i}}$ are linearly independent, the generalized holomorphic anomaly equation (\ref{generalizedBCOV00})  can be written as 
\begin{eqnarray}
\frac{\partial F^{(g_1,g_2)}}{\partial S^{jk}} = \frac{1}{2}\big{(}D_jD_k F^{(g_1,g_2-1)}+{\sum_{r_1,r_2} }^{\prime}  D_jF^{(r_1,r_2)}D_kF^{(g_1-r_1,g_2-r_2)}\big{)}
\end{eqnarray}
This equation can be integrated with respect to $S^{jk}$ and we solve for $F^{(g_1,g_2)}$ recursively as a polynomial of $S^{jk}$ with rational function coefficients, up to a rational function ambiguity. We note $S^{jk}$ is a symmetric tensor, so for the two-parameter model such as the local $\mathbb{P}^1\times \mathbb{P}^1$ model we consider here, we have 3 independent generators $S^{11}, S^{12}, S^{22}$. To carry out the polynomial formalism, we need the formula for the derivative of the propagators and the Christoffel symbol. This can be derived from the special geometry relation, see e.g. \cite{AL, HKR}. 
\begin{eqnarray}
D_iS^{jk} &=& -C_{imn}S^{jm}S^{kn}+f^{jk}_i  \\
\Gamma^k_{ij} &=& -C_{ijl}S^{kl}+\tilde{f}^k_{ij} \label{GammaS} 
\end{eqnarray}
Where the three point coupling $C_{ijk}$ are 
\begin{eqnarray}
&& C_{111}= \frac{(1-4z_2)^2-16z_1(1+z_1)}{4z_1^3\Delta},~~ 
 C_{112} \frac{16z_1^2-(1-4z_2)^2}{4z_1^2z_2\Delta} , \nonumber \\
 && C_{122} \frac{16z_2^2-(1-4z_1)^2}{4z_1z_2^2\Delta} , ~~
C_{222}= \frac{(1-4z_1)^2-16z_2(1+z_2)}{4z_2^3\Delta},
\end{eqnarray}
and the other combinations follow by symmetry. The rational functions $f,\tilde{f}$  are
\begin{eqnarray} \label{tildef}
\tilde{f}^1_{11}=-\frac{1}{z_1},~~ \tilde{f}^1_{12}= \tilde{f}^1_{21}=-\frac{1}{4z_2},~~ \tilde{f}^1_{22}=0,\nonumber\\
\tilde{f}^2_{11}=0,~~ \tilde{f}^2_{12}= \tilde{f}^2_{21}=-\frac{1}{4z_1},   ~~ \tilde{f}^2_{22}=-\frac{1}{z_2},
\end{eqnarray}
\begin{eqnarray}
&&f_1^{11}=-\frac{1}{8}z_1(1+4z_1-4z_2),  ~~  f_1^{12}= f_1^{21}= -\frac{1}{8}z_2(1+4z_1-4z_2),  \nonumber \\
&& f_1^{22}=-\frac{z_2^2}{8z_1}(1+4z_1-4z_2),\nonumber \\
&& f_2^{11}=-\frac{z_1^2}{8z_2}(1+4z_2-4z_1), ~~f_2^{12}=f_2^{21}=-\frac{1}{8}z_1(1+4z_2-4z_1), \nonumber \\
&& f_2^{22}=-\frac{1}{8}z_2(1+4z_2-4z_1)
\end{eqnarray}
Here we note that the $\tilde{f}$ are chosen so that the overdetermined equations for the propagators (\ref{GammaS}) are solvable. Under this choice of $\tilde{f}$ in (\ref{tildef}), the propagators are related and have only one independent component 
\begin{equation}
S^{ij}=\begin{pmatrix}
S(z_1,z_2) & \displaystyle\frac{z_2}{z_1}\,S(z_1,z_2) \\
\displaystyle\frac{z_2}{z_1}\,S(z_1,z_2) & \displaystyle\frac{z_2^2}{z_1^2}\,S(z_1,z_2)
       \end{pmatrix}
\end{equation}
Like the Christoffel symbol, the propagators are in general not rational functions of $z_i$. We calculate the series expansion of the propagators around the large volume point $z_1=z_2=0$ in terms of the mirror maps $Q_1, Q_2$ 
\begin{eqnarray}
S &=& \frac{Q_1^2}{2}-4Q_1^2 \left(Q_1+Q_2\right) +Q_1^2 \left(17 Q_1^2+20 Q_1Q_2 +17 Q_2^2\right)  \nonumber \\&& 
-4 Q_1^2 \left(13 Q_1^3+17 Q_1^2 Q_2 +17 Q_1 Q_2^2 +13 Q_2^3\right)  +\mathcal{O}(Q^5) 
\end{eqnarray}   

\begin{figure}
\center
\begin{picture}(0,0)%
\includegraphics{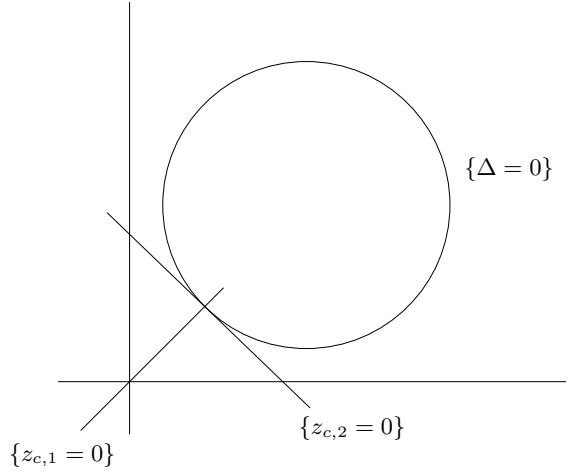}%
\end{picture}%
\setlength{\unitlength}{2072sp}%
\begingroup\makeatletter\ifx\SetFigFont\undefined%
\gdef\SetFigFont#1#2#3#4#5{%
  \reset@font\fontsize{#1}{#2pt}%
  \fontfamily{#3}\fontseries{#4}\fontshape{#5}%
  \selectfont}%
\fi\endgroup%
\begin{picture}(6672,5578)(2206,-6257)
\put(7651,-2761){\makebox(0,0)[lb]{\smash{{\SetFigFont{9}{10.8}{\rmdefault}{\mddefault}{\updefault}$\{\Delta = 0\}$}}}}
\put(5671,-5866){\makebox(0,0)[lb]{\smash{{\SetFigFont{9}{10.8}{\rmdefault}{\mddefault}{\updefault}$\{z_{c,2}=0\}$}}}}
\put(2206,-6181){\makebox(0,0)[lb]{\smash{{\SetFigFont{9}{10.8}{\rmdefault}{\mddefault}{\updefault}$\{z_{c,1}=0\}$}}}}
\end{picture}%
\caption{Conifold coordinates}
\label{conifig}
\end{figure}

The fix the rational function of $z_i$ appearing as the constant term in the integration of the holomorphic anomaly equations, we need to expand the topological strings around the conifold point of the moduli space. This is depicted in Figure \ref{conifig}. Here we choose to expand around a symmetric point $z_{c1}=z_{c2}=0$, where the coordinates are 
\begin{eqnarray}
z_{c1}= 1-\frac{z_1}{z_2},~~~  z_{c2}=1-\frac{z_2}{\frac{1}{8}-z_1}
\end{eqnarray} 

We can solve the Picard-Fuchs system of differential equations around this point, find the mirror maps 
\begin{eqnarray} \label {P1P1conimap1}
t_{c1} &=& -\log (1-z_{c1}) \\
t_{c2} &=&  z_2+\frac{1}{16} \left(2 z_1^2+8 z_1z_2 +13 z_2^2\right) +
 \frac{1}{768} \left(96 z_1^3+228 z_1^2 z_2 +240  z_1z_2^2+521 z_2^3\right)   \nonumber \\ &+&
 \frac{1}{8192} (904 z_1^4+1600  z_1^3z_2+1172  z_1^2z_2^2+1680 z_1 z_2^3+4749 z_2^4)+\mathcal{O}(z^5) \nonumber 
\end{eqnarray}
and the inverse series 
\begin{eqnarray}  \label {P1P1conimap2}
z_{c1} &=& 1- e^{-t_{c1}}   \\
z_{c2} &=& t_2-\frac{1}{16} \left(2 t_1^2+8 t_1 t_2 +13 t_2^2\right)   + \frac{1}{768} \left(48 t_1^3+312  t_1^2t_2 +696 t_1 t_2^2+493 t_2^3\right)  \nonumber \\ &-&
\frac{1}{24576}(832 t_1^4+7040  t_1^3t_2+21216 t_1^2 t_2^2+27856 t_1 t_2^3+12427 t_2^4)
+\mathcal{O}(t^5)  \nonumber
\end{eqnarray} 
The coordinate $t_{c2}$ is normal to the conifold loci, so the expansion of
the topological string amplitude 
around this point should be singular as $t_{c2}\rightarrow 0$ and  
exhibit the gap condition. To expand the topological strings 
around the conifold point $z_{c1}=z_{c2}=0$ in terms of the mirror 
maps $t_{c1}, t_{c2}$, we transform the coordinates and the propagators to the conifold coordinate
\begin{eqnarray}
S^{z_iz_j}= \frac{\partial z_i}{\partial z_{c,k}} \frac{\partial z_j}{\partial z_{c,l}} S^{z_{c,k}z_{c,l}},
\end{eqnarray}
In order to compute the series expansion of the propagator at the conifold point using (\ref{GammaS}), we also need to transform the three point Yukawa couplings,  the holomorphic ambiguity $\tilde{f}$ to the conifold coordinates. The Yukawa couplings transform as a tensor, and the  holomorphic ambiguity $\tilde{f}$ transform according the rules 
\begin{eqnarray}
\tilde{f}^{z_{c,i}}_{z_{c,j}z_{c,k}}= \frac{\partial z_{c,i}}{\partial z_l}\frac{\partial^2 z_l}{\partial z_{c,j}\partial z_{c,k}}
+ \frac{\partial z_{c,i}}{\partial z_l}\frac{\partial z_m}{\partial z_{c,j}}\frac{\partial z_n}{\partial z_{c,k}} \tilde{f} ^{z_l}_{z_mz_n}
\end{eqnarray}
We also need to calculate the Christoffel connection around the conifold point using the mirror maps (\ref{P1P1conimap1}, \ref{P1P1conimap2}), and the relation (\ref{P1P1Christoffel}). It turns out that the propagators have only one non-vanishing component around the conifold point. The three components vanish $S^{z_{c1}z_{c1}}=S^{z_{c1}z_{c2}}=S^{z_{c2}z_{c1}}=0$, and the last component is computed as 
\begin{eqnarray}
S^{z_{c2}z_{c2}} &=& \frac{t_2}{2} -\frac{1}{8} \left(4 t_1+13 t_2\right) t_2+\frac{  t_2 \left(840 t_1^2+4032 t_1 t_2+4987 t_2^2\right)
 }{1536} +\mathcal{O}(t^4) \nonumber
\end{eqnarray}
The gap conditions at the conifold point plus one more condition from the
constant map contribution at large volume point are sufficient to fix the
topological string amplitudes. Here for convenience, we fix the large volume
behavior of the topological string amplitudes to be vanishing, i.e.
$F^{(g_1,g_2)}\sim \mathcal{O}(Q)$. The constant map contribution can 
be simply recovered by adding the appropriate constant to the topological 
strings without effects on the gap condition around the conifold point.  
We found the genus two results as the followings 
\begin{eqnarray}
   F^{(0,2)} &=& \frac{5 S^3}{24 z_1^6\Delta^2 }+\frac{S^2}{48 z_1^4 \Delta^2 }  (48
   z_1^2-96  z_1z_2+40 z_1+48 z_2^2+40 z_2-13)  \nonumber \\ &&
   +\frac{S}{144 z_1^2 \Delta^2 } (384 z_1^3-384 z_1^2z_2 +80 z_1^2-384 z_1 z_2^2 +736  z_1z_2-112 z_1   \nonumber \\ &&
+384z_2^3+80 z_2^2-112 z_2+17)   +\frac{1}{1440 \Delta^2 } (2688 z_1^4+1536z_1^3 z_2 -416 z_1^3
  \nonumber \\ &&  -8448 z_2^2 z_1^2+6560 z_2 z_1^2-696 z_1^2+1536
    z_1z_2^3+6560  z_1z_2^2-2768  z_1 z_2  \nonumber \\ &&
    +258 z_1+2688 z_2^4-416 z_2^3-696 z_2^2+258 z_2-25)
    \end{eqnarray} 
\begin{eqnarray}
 F^{(1,1)} &=&
   \frac{S^2 \left(1-4 z_1-4 z_2\right)}{24 z_1^4\Delta^2}+\frac{S}{144 z_1^2\Delta^2 }  (-192 z_1^3+192  z_1^2z_2 +16 z_1^2+192z_1 z_2^2       \nonumber \\&&  -544 z_1 z_2 +28 z_1-192
   z_2^3+16 z_2^2+28 z_2-5)  +\frac{1}{720\Delta^2 } (-1408 z_1^4 \nonumber \\ && -1536 z_1^3 z_2 +736 z_1^3+5888z_1^2 z_2^2 -4320 z_1^2 z_2 -24 z_1^2  -1536
    z_1z_2^3 -4320 z_1 z_2^2  \nonumber \\ && +1328 z_1z_2  -38 z_1-1408 z_2^4+736 z_2^3-24 z_2^2-38 z_2+5) 
   \end{eqnarray}
\begin{eqnarray}
 F^{(2,0)} &=&
\frac{S}{288 z_1^2 \Delta^2}   \left(16 z_1^2+32  z_1z_2 -8 z_1+16 z_2^2-8 z_2+1\right)+\frac{1}{2880\Delta^2}( -512 z_1^4    \nonumber \\&& +9216z_1^3 z_2 +704 z_1^3-17408  z_1^2z_2^2 +2880
   z_1^2  z_2 -336 z_1^2+9216 z_1 z_2^3  +2880  z_1 z_2^2    \nonumber \\&&  -1568 z_1z_2  +68 z_1-512 z_2^4+704 z_2^3-336 z_2^2+68 z_2-5)
   \end{eqnarray}
where we have used the large volume coordinate and $S=S^{z_1z_1}$, but the
expressions are exact can be expanded around any point in the moduli space. 
We also solve the topological string amplitudes at genus three. We 
check that the expansion around large volume point $z_1=z_2=0$  
agree with the calculations (\ref{P1P1genustwo}, \ref{P1P1genusthree}) 
from the refined topological vertex. Using (\ref{generalformFmn}), the form of
the conifold discriminant (\ref{conifolddiscp1p1}) and the finiteness of the
$F^{(g_1,g_2)}$ in the large $z_{1,2}$ limit one can easily see as in~\cite{HKR}  
that the deformed model is completely integrable, i.e. all
$c^{(g_1,g_2)}_0(z_{1,2})$ are fixed by the boundary conditions.

\subsection{The refined topological string on local $\mathbb{P}^2$}
\label{P2}

The refined topological vertex formalism \cite{IKV} is not 
applicable to the local $\mathcal{O}(-3)\rightarrow \mathbb{P}^2$  
model, because it does not  give rise to a gauge theory. 

It remains an interesting problem to find a refined topological vertex 
formalism that applies directly to this model. 
Nevertheless, the homology of local $\mathbb{P}^2$ can be 
embedded into the one of another toric geometry, 
the local $\mathbb{F}_1$ model, which is the simply the blow 
up of $\mathbb{P}^2$ and geometrically engineers the $SU(2)$ 
Seiberg-Witten theory with $N_f=0$ fundamental flavor 
in four dimensions. One can calculate the 
refined topological string amplitudes of the local $\mathbb{F}_1$ model 
with the refined topological vertex formalism, and extract the BPS 
numbers $N_{j_L,j_R}$ with spins in both left and right $SU(2)$ 
subgroups of the Lorentz group. These BPS numbers determine also 
the refined BPS numbers $N_{j_L,j_R}$ of the $\mathbb{P}^2$
geometry \cite{IKV}. 

Using these refined BPS numbers provided in~\cite{IKV} we find the 
instanton part of the refined topological string amplitudes to the first few orders. The genus one and two results are
\begin{eqnarray}
F^{(0,1)}&=&\frac{Q}{4}-\frac{3 Q^2}{8}-\frac{23 Q^3}{3} + \frac{3437 Q^4}{16}+\mathcal{O}(Q^5)  \nonumber \\
F^{(1,0)}&=&\frac{7 Q}{8}-\frac{129 Q^2}{16}+\frac{589 Q^3}{6}-\frac{43009 Q^4}{32}+\mathcal{O}(Q^5)   \label{P2F10}\\
F^{(0,2)}&=&\frac{Q}{80}+\frac{3 Q^3}{20}-\frac{514 Q^4}{5}+\mathcal{O}(Q^5) \nonumber \\
F^{(1,1)}&=&\frac{11 Q}{160}-\frac{9 Q^2}{16}-\frac{1317 Q^3}{40}+\frac{72019 Q^4}{40}+\mathcal{O}(Q^5) \nonumber \\
F^{(2,0)}&=&\frac{29 Q}{640}-\frac{207 Q^2}{64}+\frac{18447 Q^3}{160}-\frac{526859 Q^4}{160}+\mathcal{O}(Q^5)  \label{P2F2}
\end{eqnarray}
And the genus three results 
\begin{eqnarray}
F^{(0,3)}&=&\frac{Q}{2016}+\frac{Q^2}{336}+\frac{Q^3}{56}+\frac{1480 Q^4}{63}+O\left(Q^5\right) \nonumber \\
F^{(1,2)}&=&\frac{127 Q}{40320}-\frac{31 Q^2}{3360}+\frac{547 Q^3}{1120}-\frac{293777 Q^4}{315}+\mathcal{O}(Q^5) \nonumber \\
F^{(2,1)}&=&\frac{143 Q}{53760}-\frac{547 Q^2}{2240}-\frac{182901 Q^3}{4480}+\frac{4107139 Q^4}{840}+\mathcal{O}(Q^5) \nonumber \\
F^{(3,0)}&=&\frac{137 Q}{322560}-\frac{7573 Q^2}{13440}+\frac{608717 Q^3}{8960}-\frac{21873839 Q^4}{5040}+\mathcal{O}(Q^5)  \label{P2F3}
\end{eqnarray}

Now we turn to the B-model calculations. We extended the approach in \cite{HKR} to solve the extended holomorphic anomaly equations. The Picard-Fuchs differential equation is well known $\mathcal{L}f=0$ where the Picard-Fuchs operator is 
\begin{equation}
\mathcal{L} = \theta^3 +3z(3\Theta+2)(3\Theta+1)\Theta 
\end{equation}
Here $\Theta=z\frac{\partial}{\partial z}$. The discriminant of the Picard-Fuchs operator is $\Delta=1+27z$. The large volume point is $z\sim 0$ and the conifold point is $z\sim -\frac{1}{27}$. The three linearly independent solutions at large volume point are the constant, a logarithmic solution and a double logarithmic solution. The K\"ahler parameter in the A-model is the logarithmic solution and its exponential $Q=e^{-T}$  is expanded as the following 
\begin{eqnarray} \label{P2mirrormap}
Q =z-6 z^2+63 z^3-866 z^4+13899 z^5+\mathcal{O}(z^6) 
\end{eqnarray}
From the double logarithmic solution we can find the three-point Yukawa coupling 
\begin{eqnarray}
C_{zzz} = -\frac{1}{3}\frac{1}{z^3(1+27z)}
\end{eqnarray}
The genus one topological amplitude $F^{(0,1)}$ is well known 
\begin{eqnarray}
F^{(0,1)} = -\frac{1}{2}\log(\frac{\partial T}{\partial z})-\frac{1}{12}\log(z^7\Delta)
\end{eqnarray}
From our general analysis we expect the refined topological string amplitude $F^{(1,0)}$ to be $1/24$ the logarithm of the discriminant, plus a piece proportional to $\log(z)$. We use the results (\ref{P2F10}) from A-model to fix this constant, and we find 
\begin{eqnarray}
F^{(1,0)}= \frac{1}{24}\log(\frac{\Delta}{z})
\end{eqnarray}
Once the constant is fixed, the large volume expansion of the above equation agrees with (\ref{P2F10}) using the mirror map (\ref{P2mirrormap}). 

To compute the higher genus refined amplitudes, we need the propagator $S^{zz}$ and its relation with the Christoffel connection. This is also fixed in \cite{HKR}
\begin{eqnarray}
\Gamma^z_{zz} &=& -C_{zzz}S^{zz}-\frac{7+216z}{6z\Delta} \nonumber \\
D_z S^{zz} &=& -C_{zzz}(S^{zz})^2-\frac{z}{12\Delta} 
\end{eqnarray}
The Christoffel symbol and the propagator can be expanded around any point in the moduli space using the corresponding mirror map around the relevant point. For example, at the large volume point $z\sim 0$, the propagator $S\equiv S^{zz}$ is 
\begin{eqnarray}
S=\frac{Q^2}{2}+15 Q^3+135 Q^4+785 Q^5+\frac{4473 Q^6}{2}+\mathcal{O}(Q^7)
\end{eqnarray}
The extended holomorphic anomaly equation of one-parameter model without the Griffiths infinitesimal invariant for genus greater than one is 
\begin{eqnarray}
\frac{\partial F^{(g_1,g_2)}}{\partial S} = \frac{1}{2}[D_z^2 F^{(g_1,g_2-1)}+(\sum_{r_1=0}^{g_1}\sum_{r_2=0}^{g_2} )^\prime D_zF^{(r_1,r_2)}D_zF^{(g_1-r_1,g_2-r_2)}]
\end{eqnarray}
where $(\sum)^\prime$ denotes the sum excludes $r_1=r_2=0$ and $r_1=g_1,r_2=g_2$. Integrating the holomorphic anomaly equation can determine the refined topological amplitude $F^{(g_1,g_2)}$ as a polynomial of $S$ whose coefficients are rational functions of $z$, up to a rational of function of the form $\frac{f(z)}{\Delta(z)^{2g_1+2g_2-2}}$, where $f(z)$ is a degree  $2g_1+2g_2-2$ polynomial. Expanding again the $F^{(g_1,g_2)}$ around the conifold point, we can fix the ambiguous rational function up to a constant, which can be further fixed by the constant map contribution at the large volume point. Since the constant map contribution can be simply recovered by adding the appropriate constant to the topological strings without effects on the gap condition around the conifold point, here for convenience we simply require $f(z)$ to a polynomial of one degree less, thus completely fix the ambiguity. The results at the first few orders are the followings  
\begin{eqnarray}
F^{(0,2)}&=& \frac{100 S^3-90 S^2 z^2+30 S z^4+3 (9 z-1) z^6}{4320 z^6 \Delta^2} \nonumber \\
F^{(1,1)}&=& \frac{10 S^2+5 S (108 z-1) z^2+2 (1-54 z) z^4}{1440 z^4 \Delta^2} \nonumber \\
F^{(2,0)}&=& \frac{10 S+(1296 z+11) z^2}{11520 z^2 \Delta^2}
\label{p2g=2}
\end{eqnarray}
The genus three results are 
\begin{eqnarray}
F^{(0,3)}&=& \frac{1}{8709120 z^{12} \Delta^4} [33600 S^6-84000 S^5 z^2-6720 S^4 (189 z-13) z^4  \nonumber \\ &&
-280 S^3 \left(17496 z^2-8964 z+173\right) z^6  +3024 S^2 \left(4941
   z^2-561 z+5\right) z^8  \nonumber \\ && +504 S \left(26244 z^3-20169 z^2+981 z-5\right) z^{10}  \nonumber \\ &&  +3 \left(-3254256 z^3+649296 z^2-18288
   z+53\right) z^{12} ]  \nonumber \\
F^{(1,2)}&=& \frac{1}{8709120 z^{10} \Delta^4}[ 8400 S^5+2520 S^4 (180 z-7) z^2+420 S^3 \left(19440 z^2-4572 z+35\right) z^4  \nonumber \\ && -42 S^2 \left(1382184 z^2-64584
   z+145\right) z^6-252 S \left(629856 z^3-221859 z^2+4977 z-5\right) z^8 \nonumber \\ && +\left(116680824 z^3-13878702 z^2+189810
   z-89\right) z^{10}]
   \nonumber \\
F^{(2,1)}&=& \frac{1}{11612160 z^8 \Delta^4}  [1120 S^4+280 S^3 (432 z-7) z^2+252 S^2 \left(36288 z^2-1788 z+5\right) z^4 \nonumber \\ && +14 S \left(22674816 z^3-3936600 z^2+38988
   z-25\right) z^6 \nonumber \\ && -\left(365316480 z^3-22651488 z^2+175608 z+253\right) z^8 ] \nonumber \\
F^{(3,0)}&=& \frac{1}{69672960 z^6 \Delta^4} [280 S^3+420 S^2 (108 z-1) z^2+42 S \left(209952 z^2-4212 z+5\right) z^4
\nonumber \\ && +\left(1167753024 z^3-29387448 z^2+355536
   z+2269\right) z^6 ]
\label{p2g=3}
\end{eqnarray}
The expansions of these exact results around the large volume point agree with 
the A-model results (\ref{P2F2}, \ref{P2F3}). Again a counting of the
parameters in the $c^{(g_1,g_2})_0$ polynomials shows similar as in~\cite{HKR}, 
that all ambiguities are fixed for the deformed 
topological string on $\mathbb{P}^2$.

\section{Conclusion and directions for further work}

We have extended the direct integration method 
developed in~\cite{HK1, HK2, HKR} to solve 
pure Seiberg-Witten theory and topological 
string theory on local Calabi-Yau spaces. We 
found a generalized holomorphic anomaly equation, 
which as we argued by comparing with the general
expansion of $F^{(g_1,g_2)}$ in terms of BPS
invariants should hold in full generality 
for the topological string on non-compact 
Calabi-Yau manifolds. We have demonstrated 
that the gap condition of the $F^{(g_1,g_2)}$ 
at the conifold provides together with regularity 
of the $F^{(g_1,g_2)}$ at other boundary divisors 
enough boundary conditions to solve these models.  
Our formalism implies that the  $F^{(g_1,g_2)}$ are 
expressible in terms of generators of a ring of 
an-holomorphic modular forms and that 
$F^{(g_1=0,g_2)}=F^{(g_1)}$ is the most 
an-holomorphic object. Our expressions, 
e.g. (\ref{p2g=2},\ref{p2g=3}) can be readily 
expanded near the $\mathbb{C}/\mathbb{Z}_3$ 
orbifold point in the local $\mathbb{P}^2$ 
moduli space using the flat coordinates 
provided in~\cite{Aganagic:2006wq} to yield 
refined orbifold Gromov-Witten invariants.

The holomorphic anomaly equation in this paper 
also applies to the pure N=2 supersymmetric 
gauge theories.  However it it seems not to 
apply to the cases with general matter. It was found, e.g. 
for the $SU(2)$ gauge theory with one fundamental 
flavor, that it has to be modified in an interesting 
way by a Griffith infinitesimal invariant~\cite{Krefl:2010fm}. 
From the point of view of direct integration, which is 
based on the fact that the  $F^{(g_1,g_2)}$ are finitely 
generated by independent generators, it  would be 
interesting to clarify the modularity property of 
this function.  

So far the versions of the holomorphic anomaly 
in this paper and in~\cite{Krefl:2010fm} are not 
derived from first principles. In the case of 
topological string case on local Calabi-Yau 
manifolds it can be argued that the $F^{(g_1,g_2)}$ 
should obey $T$-duality of the spacetime geometry, 
which in our cases is an elliptic curve. Because 
of the special properties of the quasi modular 
generator at weight 2 $E_2$, the failure of 
holomorphicity is then closely related to a 
failure of $T$-duality, which can maybe be 
explained from the space-time point of view. 

Since the $\beta$-ensemble of the matrix model was already shown to 
describe the gauge theory perturbatively~\cite{Mironov:2010ym},
it seems clear that a proof of the generalized holomorphic 
anomaly equations along the lines of~\cite{Eynard:2007hf} 
should be possible for the $N=2$ supersymmetric gauge theories, 
once the program of the paper~\cite{BKMP} is established to the deformed 
case. One expects that in this program only the recursion 
relation changes as a consequence of the $\beta$ dependent measure, 
that affects the loop equations, while the spectral curve 
stays the same. 

As was mentioned a greater challenge is to extend the 
Chern-Simons matrix model to the topological as already 
the perturbative calculation for the blown up conifold, which 
expected to related to the Chern-Simons model in the 
$\beta$-ensemble fails. It could be that a 
more general coordinate transformation, which involve 
the $\epsilon_{1/2}$ in the mirror map, is necessary
to relate the topological string on compact Calabi-Yau 
manifolds to the $\beta$-ensemble. This would be 
very useful to extend the analysis to the open 
topological string invariants and to proof the 
extended holomorphic anomaly equation as in~\cite{Eynard:2007hf}. 
Moreover it is known that the deformed $\beta$-ensemble 
calculates for $\beta=\frac{1}{2}$ $SP(N)$ 
orientifold graphs and  for $\beta=2$ $SO(N)$
orientifold graphs, which should have the 
corresponding interpretation in the topological 
string theory.  
 
We generalized the heterotic amplitude and made 
predictions for the refined topological 
invariants related to left and right $SU(2)_{R/L}$
Lefshetz action in the $K3$-fiber of a regular 
$K3$ fibered Calabi-Yau space and made a 
connection to a G\"ottsche formula for Hilbert 
schemes on $K3$ and checked that the refined 
holomorphic anomaly equation holds in this case. 

The $K3$ results could bear implications 
for the micro- and macroscopic description of 
small black holes in $K3$ fibered Calabi-Yau 
spaces.

The duality on the elliptic 
fiber of a half $K3$ relates the refined topological 
string invariants to refined topological 
invariants on the moduli space of $N=4$ 
Yang-Mills instantons on del Pezzo surfaces. Some 
predictions along these lines can be found in section 
(\ref{halfk3})

Let us finally mentioned that all our results are symmetric 
in $\epsilon_{1/2}$. That is not necessarily the 
case for general refined BPS invariants in compact
Calabi-Yau manifolds. It is also clear that in 
this case the individual refined BPS numbers are
not symplectic invariants, but  depend rather on 
the complex structure. It is nevertheless 
remarkable that a straightforward generalization 
of the formalism to the quintic yields an integer structure. 
If we extend the Ans\"atze for $F^{(1,0)}$ and $F^{(0,1)}$ 
in a natural  way to the case of the 
quintic in $\mathbb{P}^4$~\footnote{Here we use the conventions 
of~\cite{Bershadsky:1993ta} with an additional $1/2$ in 
front of $F^{(1)}=F^{(0,1)}$, which was corrected in \cite{BCOV} as it 
is essential for the higher genus calculations.}
\begin{equation} 
F^{(0,1)}=\log\left(G^{-{1/2}}_{\psi,\bar \psi} 
\exp[\frac{31}{3} K] |\psi^\frac{31}{3}(1-\psi^5)^{-\frac{1}{12}}|^2\right)
\end{equation}       
and 
\begin{equation} 
F^{(1,0)}=\log\left(\exp[\frac{31}{3}
  K]\psi^\frac{31}{3}(1-\psi^5)^\frac{1}{24}\right)\ ,
\end{equation}    
which is compatible with regularity of 
$F^{(1,0)}$ at the orbifold point, the universal 
conifold behavior and assumes that  $F^{(1,0)}$ is the 
section of the same K\"ahler line bundle then  
$F^{(0,1)}$ one gets the following integers 
\begin{equation} 
n^{(1,0)}_d= -1492, -171409, 123200314, 381613562015, \ldots
\end{equation} 
The integrality, which is nontrivial from the 
multi covering formula and the subtraction 
of the genus zero contribution, has been checked 
up to high degree d=50. It would be interesting
to understand, whether these integers count 
refined cohomologies of D-branes for some canonical choice of 
the complex structure of the quintic.

\vspace{0.2in} {\leftline {\bf Acknowledgments:}}

We thank Bernard de Wit, Babak Haghighat, Denis Klevers, 
Marco Rauch, Samson Shatasvhili, Cumrun Vafa and especially 
Marcos Mari\~no and Johannes Walcher for discussions. 
We also thank Sheldon Katz and Rahul Pandharipande for 
e-mail correspondence.     

\section{Appendix}

\subsection{$g_1+g_2=3$ results for pure $N=2$ SU(2) SYM theory}
\label{su(2)3}  

The genus 3 formulae are  
\begin{eqnarray}
F^{(0,3)} &=& \frac{1}{2916(u^2-1)^4}\big{[}5X^6-25uX^5+(50u^2+30)X^4-\frac{u}{12}(559u^2+1557)X^3 \nonumber \\
&& +\frac{1}{80}(1223u^4+13794u^2+3735)X^2-\frac{u}{40}(155u^4+3060u^2+3537)X \nonumber \\
&& +\frac{1}{3360}(236u^6+43299u^4+111078u^2+16875)\big{]} \nonumber \\
F^{(1,2)} &=&  \frac{1}{5832(u^2-1)^4}\big{[} 5uX^5-(26u^2+15)X^4+\frac{3u}{4}(75u^2+233)X^3 \nonumber \\
&& -\frac{1}{20}(1163u^4+13419u^2+3150)X^2 +\frac{u}{20}(287u^4+11751u^2+13194)X \nonumber \\
&& -\frac{3}{1120}(516u^6+52231u^4+152238u^2+23175)\big{]} \nonumber \\
F^{(2,1)} &=&  \frac{1}{5832(u^2-1)^4}\big{[}u^2X^4-u(\frac{11}{2}u^2+6)X^3+\frac{9}{20}(23u^4+184u^2+35)X^2
\nonumber \\
&& -\frac{u}{40}(53u^4+15849u^2+16182)X+\frac{1}{560}(1216u^6+93615u^4+307008u^2+45225)\big{]}
\nonumber \\
F^{(3,0)} &=&  \frac{1}{69984(u^2-1)^4}\big{[}u^3X^3-3u^2(2u^2+3)X^2-\frac{3u}{20}(u^4-1347u^2-450)X
\nonumber \\
&& -\frac{1}{140}(769u^6+87012u^4+310500u^2+43875)\big{]}   \label{genus3formulae}
\end{eqnarray} 
The dual expansions are 
\begin{eqnarray}
F_D^{(0,3)} &=& \frac{1}{1008a_D^4} -\frac{9a_D}{2^{20}}+\mathcal{O}(a_D^2) \nonumber \\
F_D^{(1,2)} &=& -\frac{41}{20160a_D^4} +\frac{15}{2^{18}}+\frac{1239a_D}{5\cdot 2^{20}}+\mathcal{O}(a_D^2) \nonumber \\
F_D^{(2,1)} &=& \frac{31}{26880 a_D^4} -\frac{117}{2^{20}}-\frac{5799a_D}{5\cdot 2^{23}}+\mathcal{O}(a_D^2) \nonumber \\
F_D^{(3,0)} &=& -\frac{31}{161280 a_D^4} -\frac{243}{2^{21}}-\frac{41607a_D}{5\cdot 2^{24}}+\mathcal{O}(a_D^2) 
\label{genus3conifold}
\end{eqnarray}
Similar to the genus two case, we observe the gap structure in the dual expansion around the conifold point with the absence of $\frac{1}{a_D}, \frac{1}{a_D^2}, \frac{1}{a_D^3}$ terms.

\subsection{$g_1+g_2=3$ results for local $\mathbb{P}^1\times\mathbb{P}^1$ }
\label{p1p13} 
 The genus three results are
 \begin{eqnarray}
F^{(0,3)} &=& -\frac{Q_1+Q_2}{3024}-\frac{4 Q_1^2+Q_1 Q_2+4 Q_2^2}{1512}-\frac{9 Q_1^3+Q_1^2Q_2 + Q_1Q_2^2+9
   Q_2^3}{1008}
  \nonumber\\ &&-\frac{1}{756} \left(16 Q_1^4+ Q_1^3Q_2+8 Q_1^2 Q_2^2+Q_1Q_2^3 +16 Q_2^4\right)
 +\mathcal{O}(Q^5)   \nonumber  \\
 F^{(1,2)} &=& -\frac{11}{30240} \left(Q_1+Q_2\right)-\frac{1}{15120}(44 Q_1^2+137 Q_1 Q_2+44 Q_2^2)\nonumber \\
 && -\frac{1}{10080}( 99Q_1^3+347  Q_1^2Q_2+347 Q_1Q_2^2 +99 Q_2^3)\nonumber \\ &&
 -\frac{1}{7560}(176 Q_1^4+641  Q_1^3Q_2+3364 Q_1^2 Q_2^2+641  Q_1Q_2^3+176 Q_2^4)  \nonumber  \\
 F^{(2,1)} &=& \frac{1}{2520} \left(Q_1+Q_2\right) +\frac{1}{2520}(8 Q_1^2-61 Q_1 Q_2+ 8Q_2^2)  \label{P1P1genusthree} \\
 && +\frac{1}{840}( 9Q_1^3-223  Q_1^2Q_2-223 Q_1Q_2^2 +9 Q_2^3)\nonumber \\ &&
 +\frac{1}{1260}(32 Q_1^4-1573  Q_1^3Q_2+28408 Q_1^2 Q_2^2-1573  Q_1Q_2^3+32 Q_2^4)   \nonumber \\
  F^{(3,0)} &=& -\frac{1}{15120} \left(Q_1+Q_2\right) -\frac{1}{7560}(4 Q_1^2+127 Q_1 Q_2+ 4Q_2^2)\nonumber \\
 && -\frac{1}{5040}( 9Q_1^3+2857  Q_1^2Q_2+2857 Q_1Q_2^2 +9 Q_2^3)          
\nonumber\\ &&
 -\frac{1}{3780}(16 Q_1^4+19531  Q_1^3Q_2+143144 Q_1^2 Q_2^2+19531  Q_1Q_2^3+16 Q_2^4)   \nonumber\end{eqnarray}

\end{document}